\documentclass[a4paper,12pt,twoside,fleqn]{article}
\usepackage[margin=1in]{geometry}
\usepackage{longtable}
\usepackage{multicol}
\usepackage{multirow}
\usepackage{amsmath}
\usepackage{amssymb}
\usepackage{bbm}
\usepackage{color}
\usepackage[T1]{fontenc}
\usepackage{pifont}
\usepackage{graphicx}
\usepackage{braket}
\usepackage[width=0.9\textwidth,footnotesize]{caption}
\usepackage{cite}  
\usepackage{subcaption}

\newcommand{\Z}[1]{\ensuremath{\mathbbm{Z}_{#1}}} 
\newcommand{\SO}[1]{\ensuremath{\mathrm{SO}(#1)}}
\newcommand{\SU}[1]{\ensuremath{\mathrm{SU}(#1)}}
\newcommand{\U}[1]{\ensuremath{\mathrm{U}(#1)}}
\newcommand{\E}[1]{\ensuremath{\mathrm{E}_{#1}}}
\newcommand{\I}{\mathrm{i}}
\newcommand{\Id}{\mathbbm{1}}

\usepackage[pdftex]{hyperref}

\hypersetup{
    pdftitle = {Flavon alignments from orbifolding: SU(5) x SU(3) model with T6/Delta(54)},
    pdfauthor = {deAnda, King, Perdomo, Vaudrevange}
}

\begin{document}

\begin{titlepage}
\begin{flushright}
\normalsize{TUM-HEP 1229/19}
\end{flushright}

\begin{center}
{\bf\Large  Flavon alignments from orbifolding: \\
$\mathbf{SU(5)\times SU(3)}$ model with $\mathbf{\mathbb{T}^6/\Delta(54)}$ 
  } \\[12mm]
Francisco~J.~de~Anda$^{\dagger}$%
\footnote{E-mail: \texttt{fran@tepaits.mx}},
Stephen~F.~King$^{\star}$%
\footnote{E-mail: \texttt{king@soton.ac.uk}},
Elena~Perdomo$^{\star}$,%
\footnote{E-mail: \texttt{e.perdomo-mendez@soton.ac.uk}}
Patrick~K.S.~Vaudrevange$^\triangle$.%
\footnote{E-mail: \texttt{patrick.vaudrevange@tum.de}}
\\[-2mm]

\end{center}
\vspace*{0.50cm}
\centerline{$^{\star}$ \it
School of Physics and Astronomy, University of Southampton,}
\centerline{\it
SO17 1BJ Southampton, United Kingdom }
\vspace*{0.2cm}
\centerline{$^{\dagger}$ \it
Tepatitl{\'a}n's Institute for Theoretical Studies, C.P. 47600, Jalisco, M{\'e}xico}
\vspace*{0.2cm}
\centerline{$^{\triangle}$ \it
Physik Department T75, Technische Universit\"at M\"unchen,}
\centerline{\it James-Franck-Stra\ss e 1, 85748 Garching, Germany}
\vspace*{1.20cm}

\begin{abstract}
{\noindent
We systematically develop the formalism necessary for ensuring that boundary conditions of flavon 
fields in extra dimensions are consistent with heterotic string theory. Having developed a set of 
consistency conditions on the boundary conditions, we explore a series of examples of orbifolds in 
various dimensions to see which ones can satisfy them. In addition we impose the further 
phenomenological requirements of having non-trivial flavon vacuum alignments and also of having 
quarks and leptons located appropriately in extra dimensions. The minimal successful case seems to 
be a 10d theory with a $\SU{3}_\mathrm{fl}$ gauged flavour symmetry, where the six-dimensional 
torus is compactified on a $\mathbb{T}^6/\Delta(54)$ orbifold. We construct a realistic $\SU{5}$ 
grand unified theory along these lines, leading to tribimaximal-reactor lepton mixing, which we 
show to be consistent with current neutrino data.
}
\end{abstract}
\end{titlepage}

\section{Introduction}

The Standard Model (SM), though remarkably successful, gives no understanding of either the origin 
of the three generations of quarks and leptons or their curious pattern of masses and mixings. 
In particular the observed very light neutrino masses and approximate tribimaximal lepton mixing 
requires new physics beyond the SM. To address some of these questions, it has been suggested that 
the three generations of quarks and leptons may be unified into a triplet of an 
$\SU{3}_\mathrm{fl}$ gauged flavour symmetry. The three generations are then analogous to 
the three colours of quarks in QCD. However, unlike QCD, the $\SU{3}_\mathrm{fl}$ gauged flavour 
symmetry must be broken spontaneously in such a way as to result in the observed quark and 
lepton masses and mixings~\cite{Ramond:1979py,Chkareuli:1981gy,Berezhiani:1983rk,Berezhiani:1985in,King:2001uz}. 

In order to spontaneously break the $\SU{3}_\mathrm{fl}$ gauged flavour symmetry, one introduces 
additional Higgs-like scalars, usually referred to as flavons. Such flavons must have certain 
vacuum alignments in $\SU{3}_\mathrm{fl}$ flavour space in order to account for the observed 
quark and lepton masses and mixings. Unfortunately, the sectors introduced in order to achieve such 
vacuum alignments are typically rather ad hoc. However, there is a top-down method for achieving 
flavon vacuum alignments coming from string theory formulated in extra dimensions. For example, 
$\E{8}\times\E{8}$ heterotic string theory in 10d can accommodate a larger gauge symmetry than the 
SM which in principle could also include a flavon sector whose vacuum alignments may be understood 
from a more robust theoretical point of view.

The approach to flavons suggested above is somewhat analogous to extra-dimensional grand unified 
theories (GUTs) compactified on orbifolds, often called ``orbifold 
GUTs''~\cite{Dienes:1998vg,Barbieri:2000vh,Kawamura:2000ev,Altarelli:2001qj,Hall:2001pg,Hebecker:2001wq,Hebecker:2001jb,Asaka:2001eh,Hebecker:2003jt,Biermann:2019amx}. 
They are typically based on $S^1/\Z{2}$ or $\mathbb{T}^2/(\Z{2}\times\Z{2})$ orbifolds, where 
Higgs doublet-triplet splitting may be achieved within the framework of extra 
dimensions~\cite{Kawamura:2000ev}. Exactly the same approach can be applied to understanding flavon 
vacuum alignments. Indeed, a 
discrete subgroup of the $\SU{3}_\mathrm{fl}$ gauged flavour symmetry may result from the 
compactification of a 6d theory down to 4d~\cite{Altarelli:2008bg,Altarelli:2006kg,Adulpravitchai:2010na,Adulpravitchai:2009id,Asaka:2001eh,Burrows:2009pi,deAnda:2018oik}. 
The connection of such field theory orbifold GUTs to string orbifolds~\cite{Dixon:1985jw,Dixon:1986jc}, 
especially the stringy origin of non-Abelian discrete flavour symmetries, has been discussed in 
ref.~\cite{Kobayashi:2006wq} (see also ref.~\cite{Beye:2014nxa}) and extended to the full string 
picture in ref.~\cite{Baur:2019kwi,Baur:2019iai}.

A recent example of the above approach to flavons in extra dimensions was discussed in 
ref.~\cite{deAnda:2018yfp}. There it was suggested that a $\SU{3}_\mathrm{fl}$ gauged flavour 
symmetry in 6d, when compactified on a torus with the orbifold $\mathbb{T}^2/(\Z{2}\times\Z{2})$ 
and supplemented by a $\Z{6}\times\Z{3}$ discrete symmetry, together with orbifold boundary 
conditions, may generate all the desired $\SU{3}_\mathrm{fl}$ breaking vacuum expectation values 
(VEVs). The analysis considered the $(0,1,-1)$ and $(1,3,-1)$ vacuum alignments (CSD3) in 
$\SU{3}_\mathrm{fl}$ flavour space of the Littlest Seesaw model~\cite{King:2013iva,King:2015dvf} 
suitable for atmospheric and solar neutrino mixing. Although this idea of using orbifold boundary 
conditions to dial the desired vacuum alignments of flavons is very attractive, it is non-trivial 
to ensure that such boundary conditions are consistent with the constraints arising from string 
theory. These constraints were ignored in~\cite{deAnda:2018yfp}, loosening the connection of that 
model with string theory.

In this paper we systematically develop the formalism necessary for ensuring that boundary 
conditions of flavon fields in extra dimensions are consistent with heterotic string theory. 
Having developed a set of consistency conditions on the boundary conditions, we explore a series of 
examples which satisfy them plus the further phenomenological requirements of yielding non-trivial 
flavon vacuum alignments and of having the SM fermions located on orbifold fixed points, so that 
their massless modes may include complete multiplets under the gauged flavour symmetry. It turns 
out to be non-trivial to satisfy all of these conditions (theoretical and phenomenological) together. 
For instance, the simple $\mathbb{T}^2/\mathbb{Z}_2$ orbifold, while allowing SM fermion matter 
localisation on fixed points, does not permit non-trivial flavon vacuum alignments, consistently 
with the formal requirements of the boundary conditions. This motivates us to go to 10d models. 
However, the simple orbifold $\mathbb{T}^6/(\Z{2}\times\Z{2})$ fares no better than the previous 
case. We find that the boundary conditions must exhibit some non-Abelian structure so that we can 
have non-trivial VEV alignments. Hence, we are led to consider 10d non-Abelian 
orbifolds~\cite{Fischer:2012qj,Konopka:2012gy,Fischer:2013qza}, where the torus is modded out, for 
example, by $S_4$ or $\Delta(54)$. The latter case is an example where we can locate the SM 
fermions on fixed points in extra dimensions. Since the $\Delta(54)$ orbifold is not so well 
studied in the literature, we develop this case in some detail, and eventually show that we can 
choose the extra dimensions in a way that we can build a complete, consistent, predictive and 
realistic model which is in principle compatible with string theory. The resulting model is capable 
of achieving the flavon vacuum alignments consistent with tri-bimaximal lepton mixing~\cite{Harrison:2002er},
which may be corrected to yield tribimaximal-reactor lepton mixing~\cite{King:2009qt}, which we 
show to be consistent with the latest neutrino data.

The layout of the paper is summarised in table~\ref{tab:orbifolds} which not only summarises the 
foregoing situation but also gives the organisation of the main body of this paper in terms of the 
section numbers~\ref{sec:Z2Orbifold},~\ref{sec:Z2xZ2Orbifold},~\ref{sec:S4Orbifold}, 
and~\ref{sec:Delta54} shown. These sections are bracketed by the extra dimensional heterotic string 
friendly formalism in section~\ref{2} and a realistic model based on the $\Delta(54)$ orbifold
in section~\ref{model}. Section~\ref{sec:conclusions} concludes the paper.

\begin{table}[!t]
\begin{center}
\begin{tabular}{|c|c|c|c|c|}
\hline
Orbifold & Flavon alignment & GUT breaking & SM matter localization & Section\\
\hline
$\mathbb{T}^2/\mathbb{Z}_2$                      & \ding{55} & \ding{52} & \ding{52} & \ref{sec:Z2Orbifold}\\
$\mathbb{T}^6/(\mathbb{Z}_2\times \mathbb{Z}_2)$ & \ding{55} & \ding{52} & \ding{55} & \ref{sec:Z2xZ2Orbifold}\\
$\mathbb{T}^6/S_4$                               & \ding{52} & \ding{52} & \ding{55} & \ref{sec:S4Orbifold}\\
$\mathbb{T}^6/\Delta(54)$                        & \ding{52} & \ding{52} & \ding{52} & \ref{sec:Delta54}\\
\hline
\end{tabular}
\caption{Orbifolds studied in this paper. We demand three necessary conditions to build a 
realistic and predictive model: a non-trivial flavon alignment, the possibility of orbifold GUT 
breaking and appropriate localizations for SM matter. Only $\mathbb{T}^6/\Delta(54)$ fulfils all of them.}
\label{tab:orbifolds}
\end{center}
\end{table}

\section{Field theory orbifolds}
\label{2}

\subsection{Constraints on the gauge embedding}
\label{sec:ConstraintsOnGaugeEmbedding}

Let us consider a field theory with $D$ extra dimensions $z$ compactified on an orbifold 
$\mathbbm{O}=\mathbbm{R}^D/S$ with space group $S$, see appendix~\ref{app:SpaceGroup} for details. 
The geometric orbifold action $z \mapsto \theta\,z + \lambda$ of a space group element 
$h = (\theta, \lambda)\in S$ (where $\theta \in P$ and the so-called point group $P$ is a finite 
subgroup of $\SO{D}$) is embedded into an action on a general field $\Phi(x,z)$ of the theory as
\begin{equation}\label{eqn:GeneralOrbifoldOfField}
\Phi(x,z) ~\stackrel{h}{\longmapsto}~ R_h\, \Phi(x,h^{-1}z)\;,
\end{equation}
where for each $h\in S$ there is an element $R_h$ of the symmetry group $\mathcal{G}$ of the 
theory. In general, the symmetry group $\mathcal{G}$ contains the internal part of the 
higher-dimensional Lorentz symmetry. For example, if $\Phi(x,z)$ is a four-dimensional scalar 
equipped with an internal vector-index we get $\Phi(x,z) \mapsto \theta\, \Phi(x,\theta^{-1}z)$ for 
$h=(\theta,0)$. In addition, $\mathcal{G}$ can contain a higher-dimensional gauged flavour symmetry 
such that $R_h$ is given by a constant gauge transformation, see for example 
ref.~\cite{Hebecker:2001jb}. We call $R_h$ the (gauge) embedding of $h$ and denote the associated 
discrete group by $R_S := \{R_h \,|\, h \in S\}$. One can apply two transformations $g$ and $h$ on 
a field $\Phi(x,z)$, either combined or one transformation after the other, i.e.
\begin{subequations}
\begin{eqnarray}
\Phi(x,z) & \stackrel{g\,h}{\longmapsto} & R_{g\,h}\, \Phi(x,(g\,h)^{-1}z)\;,\\
\Phi(x,z) & \stackrel{h}{\longmapsto}    & R_{h}\, \Phi(x,h^{-1}z)  ~\stackrel{g}{\longmapsto}~ R_{g}\,R_{h}\, \Phi(x,h^{-1}g^{-1}z) \;.
\end{eqnarray}
\end{subequations}
In both cases one has to obtain the same result. Hence, the consistency condition
\begin{equation}\label{eqn:grouphomomorphism}
R_{g\,h} ~=~ R_{g}\, R_{h}
\end{equation}
follows. Mathematically, this condition says that $R$ has to be a group homomorphism from the space 
group $S$ into the (gauge) group $\mathcal{G}$. It is easy to obtain some immediate consequences of 
eq.~\eqref{eqn:grouphomomorphism}, e.g
\begin{equation}\label{eqn:grouphomomorphismConsq1}
R_{(\Id, 0 )} ~=~ \Id \quad , \quad R_{g^{-1}} ~=~ \left(R_g\right)^{-1} \quad , \quad R_{g\,h\,g^{-1}} ~=~ R_g\,R_h\,\left(R_g\right)^{-1}\;,
\end{equation}
for $g,h \in S$, and
\begin{equation}\label{eqn:grouphomomorphismConsq2}
R_{e_i}\,R_{e_j} ~=~ R_{e_j}\,R_{e_i} \;\; , \;\; R_{\theta e_i} ~=~ R_{(\theta, \lambda)}\, R_{e_i}\, \left(R_{(\theta, \lambda)}\right)^{-1} \;\; , \;\; \left(R_{(\theta, \lambda)}\right)^N =~ \Id\;,
\end{equation}
for a space group element $(\theta, \lambda) \in S$ of order $N$, i.e.\ 
$(\theta, \lambda)^N = (\Id,0)$, $i,j\in\{1,\ldots,D\}$, and we have defined the embedding of a pure 
translation as $R_{e_i} = R_{(\Id, e_i)}$, see also section 2 of ref.~\cite{GrootNibbelink:2003gd}. 
Hence, the choices for the embedding $R$ are strongly constrained as we will also see in more 
detail in the examples of the next sections.

\subsection{Standard embedding}
\label{sec:StandardEmbedding}

There exists a simple solution to the group homomorphism condition~\eqref{eqn:grouphomomorphism}: 
the so-called standard embedding. In this case, (ignoring for a moment the higher-dimen\-sional 
Lorentz symmetry for simplicity) one chooses a gauge group $\mathcal{G}$ such that the point group 
$P$ is a discrete subgroup of it. Furthermore, for supersymmetric orbifolds in $D=6$ dimensions we 
have $P \subset \SU{3}$ such that there exists a globally defined constant spinor on the orbifold 
$\mathbbm{O}$~\cite{Dixon:1985jw,Dixon:1986jc}. Hence, in complex coordinates each point group 
element $\theta \in P$ is given by a $3 \times 3$ unitary matrix (with $\mathrm{det}(\theta)=1$) 
and the choice
\begin{equation}\label{eqn:StandardEmbedding}
R_{h} ~=~ \theta \qquad\mathrm{for}\qquad h ~=~ (\theta, \lambda) ~\in~ S\;,
\end{equation}
trivially satisfies the group homomorphism condition~\eqref{eqn:grouphomomorphism} for 
$\mathcal{G}=\SU{3}_\mathrm{fl}$. This $\SU{3}$ gauge symmetry can naturally be identified with a 
flavour symmetry, hence the subscript ``fl''. In other words, the geometrical element 
$h=(\theta, \lambda)$ that acts on the (complex) extra-dimensional coordinates 
$z \in \mathbbm{C}^3$ as $z \mapsto \theta\,z + \lambda$ is accompanied by an identical gauge 
transformation $R_{h}=\theta$ in $\SU{3}_\mathrm{fl}$ flavour space, in detail, 
$\Phi \mapsto \theta\,\Phi$ if $\Phi$ is a triplet of $\SU{3}_\mathrm{fl}$. Note that 
eq.~\eqref{eqn:StandardEmbedding} implies for instance that $R_{e_i} = \Id$ for $i\in\{1,\ldots,6\}$, 
i.e.\ there are no Wilson lines along the torus-directions~\cite{Hebecker:2001jb} in the case of 
standard embedding, and the discrete gauge embedding group is isomorphic to the point group, 
$R_S \cong P$.

As a final remark, in addition to the condition~\eqref{eqn:grouphomomorphism} on the (gauge) 
embedding, string theory orbifolds are constrained by world-sheet modular invariance of the 
one-loop string partition function~\cite{Dixon:1986jc}. However, modular invariance is 
automatically satisfied in the case of standard embedding. Thus, the standard embedding 
$R_S \cong P$ in our field theory discussion fulfils all necessary conditions for a full string 
completion.

\subsection{Orbifold boundary conditions}
\label{sec:OrbifoldBoundaryConditions}

Next, we discuss the various origins of fields $\Phi(x,z)$ on an orbifold $\mathbbm{O}$ and the 
conditions to make these fields well-defined within the orbifold construction. This discussion is 
crucial in order to understand the orbifold-alignment of flavon VEVs in flavour space.

For each space group element $g\in S$ that has a non-trivial fixed point set 
\begin{equation}
\mathrm{F}_{g} ~=~ \{z \in \mathbbm{R}^D ~|~ g\,z = z\} ~\neq~ \emptyset
\end{equation}
one can define a field $\Phi_{g}(x,z)$ that is localized on $\mathrm{F}_{g}$, i.e.
\begin{equation}\label{eqn:LocalizedField}
\Phi_{g}(x,z) ~=~ 0 \quad\mathrm{if}\quad z ~\not\in~ \mathrm{F}_{g}\;.
\end{equation}
This localized field $\Phi_{g}(x,z)$ transforms in some representation of $\mathcal{G}$. For 
example, for a flavour symmetry $\mathcal{G}=\SU{3}_\mathrm{fl}$ we will mostly assume that 
$\Phi_{g}(x,z)$ is either a singlet or a triplet of $\SU{3}_\mathrm{fl}$.

In the language of string theory, a field $\Phi_{g}(x,z)$ with $g \neq \Id$ corresponds to a 
so-called twisted string localized at the fixed point set $\mathrm{F}_{g}$ of the so-called 
constructing element $g \in S$. In contrast, a field $\Phi_{g}(x,z)$ with $g = \Id$ has a trivial 
fixed point set $\mathrm{F}_{g}=\mathbbm{R}^D$ and, hence, lives in the full bulk $\mathbbm{O}$ of 
the extra dimensions. 

Importantly, a field $\Phi_{g}(x,z)$ from eq.~\eqref{eqn:LocalizedField} is in general not yet 
invariant under the orbifold action eq.~\eqref{eqn:GeneralOrbifoldOfField}, i.e.\ it transforms as
\begin{equation}\label{eqn:GeneralOrbifoldOfField2}
\Phi_g(x,z) ~\stackrel{h}{\longmapsto}~ R_h\, \Phi_g(x,h^{-1}z)\;,
\end{equation}
for $h \in S$. Let us analyze eq.~\eqref{eqn:GeneralOrbifoldOfField2} in more detail. While 
the field $\Phi_g(x,z)$ on the left-hand side is localized at $z \in \mathrm{F}_g$, the field 
$\Phi_g(x,h^{-1}z)$ on the right-hand side is localized at $h^{-1}z \in \mathrm{F}_g$ which is 
equivalent to $z \in \mathrm{F}_{h\,g\,h^{-1}}$. However, the only field localized on the fixed point 
set $\mathrm{F}_{h\,g\,h^{-1}}$ is $\Phi_{h\,g\,h^{-1}}(x,z)$. Consequently, the fields 
$\Phi_g(x,h^{-1}z)$ and $\Phi_{h\,g\,h^{-1}}(x,z)$ have to be related,
\begin{equation}\label{eqn:IdentificationOfTwistedStrings}
\Phi_g(x,h^{-1}z) ~\sim~ \Phi_{h\,g\,h^{-1}}(x,z)\;.
\end{equation}
By definition, the space group element $h\,g\,h^{-1}$ belongs to the conjugacy class $[g]$. Thus, 
fields from the same conjugacy class have to be identical up to some proportionality factors, i.e.\ 
up to a symmetry transformation from $\mathcal{G}$.

A special case appears if $h\,g\,h^{-1} = g$ for $h \neq g$, i.e.\ in the case when $g$ and $h$ 
commute. Hence, we define the set of commuting elements (the so-called centralizer $C_g$) of the 
constructing element $g$ as
\begin{equation}
C_g ~:=~ \{ h ~\in~ S ~|~ g\, h ~=~ h\, g\} ~\subset~ S\;.
\end{equation}
Now, we have to distinguish between two cases, depending on whether $g$ and $h$ commute or not:
The case $h \not\in C_g$ is not of great importance to us and is therefore relegated to 
appendix~\ref{app:OrbifoldInvariantFields}. On the other hand, if $h \in C_g$ we get 
$\Phi_{h\,g\,h^{-1}}(x,z) = \Phi_g(x,z)$. Then, eqs.~\eqref{eqn:GeneralOrbifoldOfField2} 
and~\eqref{eqn:IdentificationOfTwistedStrings} yield a non-trivial boundary condition
\begin{equation}\label{eqn:OrbifoldBC}
\Phi_g(x,z) ~\stackrel{h}{\longmapsto}~ R_h\, \Phi_g(x,h^{-1}z) ~=~ \Phi_g(x,z) \qquad\mathrm{for}\qquad h ~\in~ C_g \quad\mathrm{and}\quad h ~\neq~ g\;.
\end{equation}
Since $z$ and $h^{-1}z$ are identified on the orbifold $\mathbbm{O}$, this boundary condition 
ensures that the field $\Phi_g$ evaluated at identified points has a unique value up to a symmetry 
transformation with $R_h \in \mathcal{G}$. The orbifold boundary condition using the constructing 
element $h=g \in C_g$ itself is special, since
\begin{subequations}
\begin{eqnarray}
\Phi_{g}(x,g^{-1}z) & = & \Phi_{g}(x,z) ~=~ 0          \ \ \mathrm{if}\ z ~\not\in~ \mathrm{F}_{g} \;,\\
\Phi_{g}(x,g^{-1}z) & = & \Phi_{g}(x,z) \phantom{~=~ 0}\ \ \mathrm{if}\ z ~\in~ \mathrm{F}_{g}     \;.
\end{eqnarray}
\end{subequations}
Hence, $\Phi_{g}(x,g^{-1}z) = \Phi_{g}(x,z)$ for all $z\in \mathbbm{R}^D$ and the 
transformation~\eqref{eqn:GeneralOrbifoldOfField2} reads 
\begin{equation}\label{eqn:OrbifoldBCwithConstrElement}
\Phi_g(x,z) ~\stackrel{g}{\longmapsto}~ R_g\, \Phi_g(x,g^{-1}z) ~=~ R_g\, \Phi_g(x,z)\;,
\end{equation}
while the relation~\eqref{eqn:IdentificationOfTwistedStrings} becomes trivial for $h=g$. In more 
detail, for $h=g$ we get $\Phi_g(x,h^{-1}z) = \Phi_g(x,z)$ on the left-hand side and 
$\Phi_{h\,g\,h^{-1}}(x,z) = \Phi_{g}(x,z)$ on the right-hand side of 
eq.~\eqref{eqn:IdentificationOfTwistedStrings}. From the string construction, we know that a 
string with constructing element $g \in S$ survives the orbifold projection at least under the 
action of $g \in C_g$, see for example appendix A.5 in ref.~\cite{Vaudrevange:2008sm}. Hence, 
eq.~\eqref{eqn:OrbifoldBCwithConstrElement} imposes the boundary condition 
$R_g\, \Phi_g(x,z) = \Phi_g(x,z)$ on the localized field $\Phi_g(x,z)$.

In summary, in order to build an orbifold-invariant field $\Phi_g(x,z)$ that is localized on the 
fixed point set $\mathrm{F}_g$ we have to impose a non-trivial boundary condition~\eqref{eqn:OrbifoldBC} 
for each space group element $h \in C_g$ that commutes with the constructing element $g \in S$ of 
the localized field $\Phi_g(x,z)$.

Let us briefly discuss the trivial example with constructing element $g=\Id \in S$. In this case, 
the fixed point set is given by $\mathrm{F}_\Id = \mathbbm{R}^D$ and the field $\Phi_\Id(x,z)$ 
lives in the full bulk $\mathbbm{O}$ of the extra dimensions. Furthermore, the centralizer $C_\Id$ 
equals the full space group, i.e.\ $C_\Id = S$, and we have to impose boundary 
conditions~\eqref{eqn:OrbifoldBC} for all elements $h \in S$, i.e.\ for all generators of $S$.

\subsection{VEV alignment from orbifold boundary conditions}
\label{sec:VEValignment}

Our main focus is to interpret $\mathcal{G}$ as a gauged flavour group in extra dimensions and 
the field $\Phi_g(x,z)$ as a flavon. Therefore, we separate the higher-dimensional Lorentz symmetry from 
$\mathcal{G}$ and consider $\mathcal{G}$ as a pure gauge symmetry. Then, orbifold boundary 
conditions~\eqref{eqn:OrbifoldBC} break the flavour symmetry $\mathcal{G}$ and simultaneously 
align the vacuum expectation values of the flavons, as we discuss next.

Consider a field $\Phi_g(x,z)$ localized at $z\in\mathrm{F}_g$ with constructing element $g \in S$. 
We take an element $h = (\theta_h, \lambda_h) \in C_g$, where the order of 
$\theta_h$ is denoted by $N_h$, i.e.\ $\left(\theta_h\right)^{N_h}=\Id$. After choosing 
$\ell_h \in \{0,\ldots, N_h-1\}$, the field $\Phi_g(x,z)$ has to satisfy the boundary 
condition~\eqref{eqn:OrbifoldBC},
\begin{equation}\label{eq:OrbifoldBCWithLorentz}
\Phi_g(x,z) ~\stackrel{h}{\longmapsto}~ \exp\left(\frac{2\pi\I\, \ell_h}{N_h}\right)\,  R_h\, \Phi_g(x,h^{-1}z) ~=~ \Phi_g(x,z)\;.
\end{equation}
The $\ell_h$-dependent phase originates from diagonalizing the higher-dimensional Lorentz 
transformation, which is possible for all $h \in C_g$ simultaneously if the centralizer 
$C_g$ is Abelian.

We denote the four-dimensional zero mode of the field $\Phi_g(x,z)$ by $\Phi_g(x)$. Then, the 
orbifold boundary condition~\eqref{eq:OrbifoldBCWithLorentz} evaluated at the vacuum expectation 
value of the zero mode $\braket{\Phi_g}$ reads
\begin{equation}\label{eqn:OrbifoldVEVBC}
\braket{\Phi_g} ~\stackrel{h}{\longmapsto}~ \exp\left(\frac{2\pi\I\, \ell_h}{N_h}\right)\, R_h\,\braket{\Phi_g} ~=~ \braket{\Phi_g}\;,
\end{equation}
for all elements $h \in C_g$ from the centralizer of the constructing element $g$. 
This condition can align the VEV of a localized field to a specific direction in flavour space. In 
other words, the VEV of a flavon located in $\mathrm{F}_g$ must be an eigenvector of the matrices 
$R_h$ with an $\ell_h$-dependent phase as eigenvalue. However, the magnitude of the VEV cannot be 
constrained by orbifold boundary conditions.

For example, take an $\SU{3}_\mathrm{fl}$ flavour group and a triplet flavon $\Phi_g$ with 
constructing element $g \in S$ and assume that the centralizer contains an element $h \in C_g$ with 
$h^3 = \Id$ such that $(R_h)^3 = \Id$. By choosing a special embedding $R_h \in \SU{3}_\mathrm{fl}$ 
and $\ell_h=0$, the flavon VEV $\braket{\Phi_g}$ is aligned according to
\begin{equation}\label{eqn:ExampleVEV}
R_h ~=~ \begin{pmatrix} 0&1&0\\0&0&1\\1&0&0\end{pmatrix} \qquad\Rightarrow\qquad \braket{\Phi_g} ~\propto~ \begin{pmatrix}1\\1\\1\end{pmatrix}\;.
\end{equation}

However, one can always choose a basis in flavour space such that a given embedding matrix $R_h$ 
becomes diagonal and the VEV aligns, for example, into the first component of $\braket{\Phi_g}$. To 
avoid this rather trivial situation, one has to ensure that the discrete embedding group 
$R_S \subset \mathcal{G}$ is non-Abelian such that one cannot diagonalize all elements 
simultaneously. This is the key observation towards successful flavon alignment.

\section[Flavour from a T2/Z2 orbifold]{\boldmath Flavour from a $\mathbb{T}^2/\Z{2}$ orbifold\unboldmath}
\label{sec:Z2Orbifold}

\begin{figure}[!t]
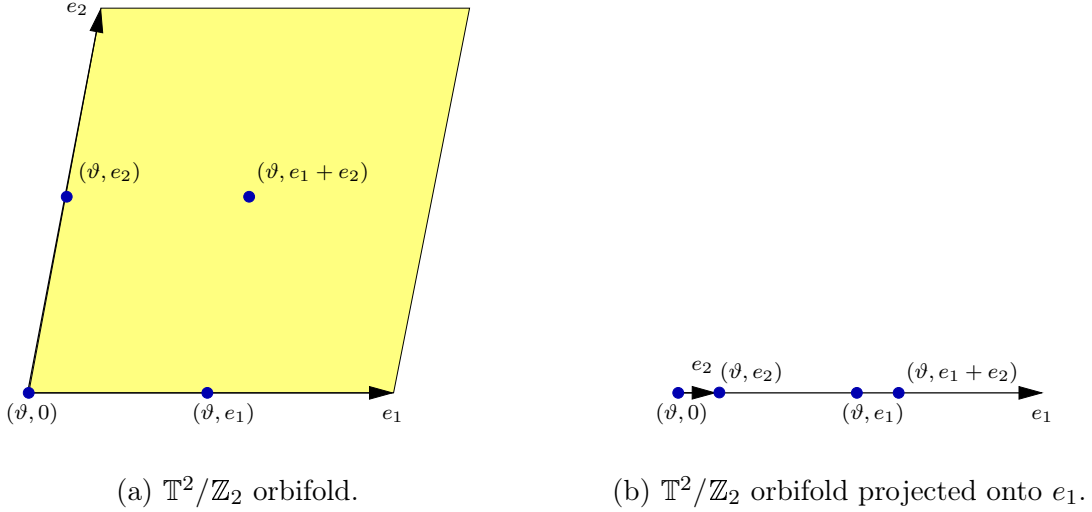

\begin{subfigure}[b]{0.5\textwidth}
\begin{center} \input{Z2Orbifoldt.pdf} \end{center}
\caption{$\mathbb{T}^2/\Z{2}$ orbifold.}
\label{step1er}
\end{subfigure}
\begin{subfigure}[b]{0.5\textwidth}
\begin{center}\input{Z2OrbifoldProjectedt.pdf}\end{center}
\caption{$\mathbb{T}^2/\Z{2}$ orbifold projected onto $e_1$.}
\label{step2er}
\end{subfigure}
\caption{(a) The two-torus $\mathbb{T}^2$ is spanned by the vectors 
$e_1$ and $e_2$ and its fundamental domain is highlighted in yellow. The $\Z{2}$ orbifold action 
$\vartheta$ maps $z \mapsto-\,z$ resulting in four fixed points labeled by their constructing 
elements. (b) The same orbifold projected onto one dimension given by the $e_1$ direction (as a 
preparation for six-dimensional orbifolds projected onto three dimensions, see e.g.\ 
figure~\ref{fig:T6Z2Z2Orbifold}).}
\label{fig:T2Z2Orbifold}
\end{figure}

Let us begin with an easy example with $D=2$ extra dimensions, parametrized in complex coordinates 
by $z \in \mathbbm{C}$. We choose a general two-torus $\mathbb{T}^2$ spanned by two vectors $e_1$ 
and $e_2$. Then, $z$ is compactified on a $\mathbb{T}^2/\Z{2}$ orbifold, i.e.\ with point group 
$P \cong \Z{2}$, where the $\Z{2}$ orbifold action $\vartheta$ is generated by $z \mapsto-z$. This 
orbifold has four inequivalent fixed points $\mathrm{F}_g$, 
\begin{equation}\label{eq:FixedPointsZ2}
\mathrm{F}_g ~\in~ \left\{\Big\{0\Big\},\ \ \Big\{\frac{e_1}{2}\Big\},\ \ \Big\{\frac{e_2}{2}\Big\},\ \ \Big\{\frac{e_1+e_2}{2}\Big\} \right\}\;,
\end{equation}
corresponding to the constructing elements $g$,
\begin{equation}\label{eq:ConstructingElementsZ2}
g ~\in~ \{(\vartheta, 0) \;,\; (\vartheta, e_1) \;,\; (\vartheta, e_2) \;,\; (\vartheta, e_1+e_2)\}\;,
\end{equation}
respectively, see figure~\ref{fig:T2Z2Orbifold}.

To define a field theory on this $\Z{2}$ orbifold, we have to choose a (gauge) embedding 
$R_{\vartheta}$, $R_{e_1}$, and $R_{e_2}$, for each generator of the $\Z{2}$ space group $S$, i.e.\ 
for each $(\vartheta, 0)$, $(\Id, e_1)$, and $(\Id, e_2) \in S$. We have to ensure that this embedding 
satisfies the following conditions, obtained from eq.~\eqref{eqn:grouphomomorphism}, 
\begin{equation}\label{eqn:Z2Conditions}
\left(R_{\vartheta}\right)^2 ~=~ \Id \quad,\quad R_{e_1}\, R_{e_2} ~=~  R_{e_2}\,R_{e_1} \quad\mathrm{and}\quad R_{\vartheta}\,R_{e_i}\,R_{\vartheta} ~=~ \left(R_{e_i}\right)^{-1}\;,
\end{equation}
where $i\in\{1,2\}$, see also eqs.~\eqref{eqn:grouphomomorphismConsq1} 
and~\eqref{eqn:grouphomomorphismConsq2}. 

\subsection{Example of a trivial VEV alignment}

A trivial solution to eq.~\eqref{eqn:Z2Conditions} is given by the choice of a gauged flavour 
symmetry $\mathcal{G} = \U{2}$ and
\begin{equation}\label{eq:Z2StandardEmbedding}
R_{\vartheta} ~=~ \begin{pmatrix} 1&0\\0&-1\end{pmatrix} \quad \mathrm{and} \quad R_{e_i} ~=~ \Id\;.
\end{equation}
In this case, the embedding group $R_S \cong \Z{2}$ is Abelian and, consequently, VEVs can only be 
aligned trivially in flavour space. In detail, take a doublet flavon $\Phi_g(x,z)$ from the bulk 
(i.e.\ with trivial constructing element $g=(\Id,0) \in S$ and centralizer $C_g = S$). Then, its 
VEV can only be aligned according to
\begin{equation}\label{eq:Z2TrivialVEVAlignment}
\braket{\Phi_{g, +}} ~\propto~ \begin{pmatrix}1\\0\end{pmatrix} \quad\mathrm{or}\quad \braket{\Phi_{g, -}} ~\propto~ \begin{pmatrix}0\\1\end{pmatrix}\;,
\end{equation}
originating from the boundary conditions
\begin{equation}
R_{\vartheta} \braket{\Phi_{g, \pm}} ~=~ \pm\braket{\Phi_{g, \pm}} \quad\mathrm{and}\quad R_{e_i} \braket{\Phi_{g, \pm}} ~=~ \braket{\Phi_{g, \pm}}\;,
\end{equation}
where the extra $\pm$ sign in the $\vartheta$-boundary condition is motivated from the 
transformation properties of $\Phi_g(x,z)$ under higher-dimensional Lorentz symmetry, see 
section~\ref{sec:VEValignment}.

In order to avoid this trivial VEV alignment eq.~\eqref{eq:Z2TrivialVEVAlignment} in the case of 
this $\Z{2}$ orbifold, we have to choose non-trivial Wilson lines $R_{e_i} \neq \Id$ for some 
$i\in\{1,2\}$. However, choosing $R_{e_i} \neq \Id$ with $(R_{e_i})^2 = \Id$ in 
eq.~\eqref{eqn:Z2Conditions} would also result in an Abelian embedding group\footnote{This 
choice corresponds to the Abelianization $R_S \cong S/[S,S]$ of the space group $S$, see 
e.g.~\cite{Blaszczyk:2012,Blaszczyk:2015ams,Ramos-Sanchez:2018edc}.} and, consequently, would 
yield a trivial VEV alignment.

\subsection{Example of a non-trivial VEV alignment}
\label{sec:t2vev}

Let us give a non-trivial example of matrices $R_\vartheta$, $R_{e_1}$ and $R_{e_2}$ for a gauged 
flavour symmetry $\mathcal{G}=\U{3}$, where $(R_{e_i})^2 \neq \Id$ for some translations $e_i$.

Since $(\vartheta)^2  = \Id$ the gauge embedding $R$ has to satisfy $(R_\vartheta)^2 = \Id$. We 
might choose
{\begin{equation}
R_\vartheta ~=~ T_{13} \quad\mathrm{where}\quad T_{13}~=~\begin{pmatrix} 0 & 0&-1 \\ 0&1&0 \\-1&0&0 \end{pmatrix} \quad\mathrm{and}\quad \left(T_{13}\right)^2 ~=~ \Id\;.
\end{equation}
Furthermore, we have to satisfy condition~\eqref{eqn:Z2Conditions}, i.e.\ 
$(R_\vartheta R_{e_i})^2=\Id$ for $i\in\{1,2\}$. To do so, we are left with the freedom to choose 
$R_{e_i}$ for $i\in\{1,2\}$ subject to the previous condition. One can check that a solution is given by
\begin{equation}
R_{e_1} ~=~ SU \, T_{13} \quad\mathrm{and}\quad R_{e_2} ~=~ T_{13}\, SU ~=~ \left(R_{e_1}\right)^{-1}\;,
\end{equation}
using the matrices
\begin{equation}
S ~=~ \frac{1}{3}\begin{pmatrix} -1 & 2&2 \\ 2&-1&2 \\2&2&-1 \end{pmatrix} \quad\mathrm{and}\quad U ~=~ \begin{pmatrix} 1&0&0\\0&0& 1 \\ 0&1&0 \end{pmatrix}\;,
\end{equation}
where $(SU)^2 = \Id$. These matrices are chosen since they are part of a specific representation of 
the discrete group $S_4$, which is known for generating predictive flavour 
structures~\cite{deAnda:2018yfp,King:2011zj}.

Importantly, for this choice the discrete gauge embedding group $R_S$ is non-Abelian and, 
furthermore, the matrices $R_{e_i}$ have infinite order, i.e.\ for all $N_i \in\mathbbm{N}$ we have
\begin{equation}
(R_{e_i})^{N_i} ~\neq~ \Id\;.
\end{equation}
Hence, $R_S$ is not a finite group, see ref.~\cite{Barbieri:2001dm} and, furthermore, 
ref.~\cite{Hebecker:2003jt} for a discussion on rank reduction in the case when the discrete gauge 
embedding group $R_S$ is non-Abelian.

Now, consider a triplet flavon $\Phi_{g}(x,z)$ with constructing element $g=(\Id,0) \in S$. It is 
localized in the bulk $\mathbbm{O}$ of the extra-dimensional space. Then, the corresponding 
centralizer $C_{g}$ can be generated by
\begin{equation}
C_{g} ~=~ \langle (\vartheta, 0) \;,\; (\Id, e_1) \;,\; (\Id, e_2) \rangle \qquad\mathrm{for}\qquad g =~ (\Id,0)\;.
\end{equation}
The flavon is subject to boundary conditions~\eqref{eqn:OrbifoldVEVBC} which result in the 
following non-trivial conditions on the VEV (choosing $\ell_h=0$ for $h=(\vartheta,0)$)
\begin{equation}
R_\vartheta\, \braket{\Phi_{g}} ~=~ \braket{\Phi_{g}} \;\;\;,\;\;\; R_{e_1}\, \braket{\Phi_{g}} ~=~ \braket{\Phi_{g}} \;\;\;,\;\;\; R_{e_2}\, \braket{\Phi_{g}} ~=~ \braket{\Phi_{g}}\;.
\end{equation}
The solution is given by a fixed VEV alignment
\begin{equation}\label{eq:VEVAlignmentForCSD3}
\braket{\Phi_{g}} ~\propto~ \begin{pmatrix}1\\3\\-1\end{pmatrix}\;,
\end{equation}
where the magnitude of the VEV cannot be determined by orbifold boundary conditions.

We can try to locate another flavon $\tilde\Phi_{g}$ in the bulk with a different $\Z{2}$ phase, 
explicitly \mbox{$R_\vartheta \braket{\tilde\Phi_{g}} = -\braket{\tilde\Phi_{g}}$}, to obtain a 
different flavon alignment. However, it turns out that $\braket{\tilde\Phi_{g}} = 0$. In other 
words, $\tilde\Phi_{g}$ is projected out by the orbifold action in this case.

We have achieved the flavon alignment~\eqref{eq:VEVAlignmentForCSD3}, which is necessary for the 
CSD3 setup~\cite{King:2013iva}. This is highly predictive for the lepton sector and usually 
complicated to obtain through a vacuum alignment 
superpotential~\cite{King:2016yvg,Bjorkeroth:2015ora,Bjorkeroth:2015uou,King:2013iva,Bjorkeroth:2017ybg,deAnda:2018oik}. 
However, it is not enough by itself. We have shown that there are no other alignments we can obtain 
through boundary conditions in this setting. Consequently, after a brief discussion on GUT breaking 
and the localization of SM matter in the following, we will go to higher-dimensional orbifolds with 
larger point groups to align various flavons simultaneously.

\subsection{GUT breaking}
\label{sec:GUTZ2}

We assume that the extra-dimensional theory before orbifolding contains a GUT gauge symmetry in 
addition to the gauged flavour symmetry $\mathcal{G}$, where we will choose 
$\mathcal{G} = \SU{3}_\mathrm{fl}$ or $\U{3}_\mathrm{fl}$ as our prime examples. Then, the full 
gauge symmetry in extra dimensions is given by
\begin{equation}
\SU{5}_\mathrm{GUT} \times \mathcal{G} \qquad\mathrm{or}\qquad \SO{10}_\mathrm{GUT} \times \mathcal{G}\;.
\end{equation}
Both, the GUT group and the flavour group, have to be broken by orbifold boundary conditions, 
determined by the gauge embedding $R$. In this specific $\mathbb{T}^2/\Z{2}$ orbifold 
the GUT-breaking boundary conditions can be chosen to correspond to any of the translations 
$R_{e_i}$, while the flavor symmetry can be broken by the gauge embedding $R_{\vartheta}$ of the 
orbifold twist $\vartheta$.

If the symmetry is $\SU{5}_\mathrm{GUT}$ the GUT-breaking boundary condition can be chosen as
\begin{equation}
P_{\SU{5}} ~=~ \mathrm{diag}(1,1,1,-1,-1) \otimes \Id_\mathrm{flavour}\;,
\end{equation}
which breaks $\SU{5}_\mathrm{GUT} \to \SU{3}_\mathrm{C}\times\SU{2}_\mathrm{L}\times\U{1}_Y$. Since 
$P_{\SU{5}}$ does not act on the flavour symmetry $\mathcal{G}$, it commutes with all flavour 
breaking conditions and, hence, is consistent with eqs.~\eqref{eqn:Z2Conditions} (using for example 
$R_{e_1} = P_{\SU{5}}$ and $(P_{\SU{5}})^2 = \Id$).

In the case of an $\SO{10}_\mathrm{GUT}$ symmetry, $\SO{10}$ can be broken by two independent 
$\Z{2}$ boundary conditions~\cite{Asaka:2001eh}
\begin{equation}
P_\mathrm{GG} ~=~ \mathrm{diag}(1,1,1,1,1)\otimes\sigma_2\otimes \Id_\mathrm{flavour}\;,\ \ \ P_\mathrm{PS} ~=~ \mathrm{diag}(1,1,1,-1,-1)\otimes \sigma_0\otimes \Id_\mathrm{flavour}\;,
\end{equation}
where $\sigma_i$ are the Pauli matrices. Each condition separately breaks $\SO{10}_\mathrm{GUT}$ 
according to
\begin{equation}
P_\mathrm{GG}: \SO{10}_\mathrm{GUT} ~\to~ \SU{5}\times \U{1}, \ \ \ P_\mathrm{PS}: \SO{10}_\mathrm{GUT} ~\to~ \SU{4}\times \SU{2}\times \SU{2},
\end{equation}
while together they break $\SO{10}_\mathrm{GUT} \to \SU{3}_\mathrm{C}\times\SU{2}_\mathrm{L}\times\U{1}_Y\times\U{1}$, 
see e.g.~\cite{Adulpravitchai:2010na}. The two boundary conditions $P_\mathrm{GG}$ and 
$P_\mathrm{PS}$ commute with each other, as well as with any flavour-breaking condition. Thus, we 
can choose each one to be one of the $R_{e_i}$ consistently.

\subsection{SM fermion localization}
\label{sec:SMMatterLocalizationZ2}

The SM matter fields, as any other field in this $\mathbb{T}^2/\Z{2}$ orbifold, must 
be located somewhere in extra dimensions: i) either on a fixed point set $\mathrm{F}_g$ for 
$g \neq \Id$, being points in compactified dimensions, see eq.~\eqref{eq:FixedPointsZ2}, or ii) in 
the two-dimensional bulk $\mathbbm{O}$. In principle, localized fields feel boundary 
conditions~\eqref{eqn:OrbifoldVEVBC} with respect to their centralizers $C_g$. Consequently, some 
zero modes of SM matter fields can be projected out by the orbifold -- depending on the respective 
centralizers $C_g$. 

First, consider a localized field $\Phi_g$ with constructing element $g$ from 
eq.~\eqref{eq:ConstructingElementsZ2}. In this case, the centralizer $C_g = \{\Id, g \}$ is 
trivial as discussed in section~\ref{sec:OrbifoldBoundaryConditions}. Consequently, a localized 
field $\Phi_g$ in the $\mathbb{T}^2/\Z{2}$ orbifold is not subject to orbifold boundary conditions 
and the four-dimensional zero mode $\Phi_g(x)$ is not projected out. Therefore, SM matter fermions 
from localized fields appear in complete GUT multiplets. This is the field-theoretical analogue 
to string-theoretical local GUTs with complete SM generations~\cite{Forste:2004ie,Buchmuller:2004hv,Lebedev:2006kn}.

In contrast, a bulk field is subject to non-trivial orbifold boundary conditions, especially to 
those that induce GUT breaking. Let us discuss the consequences of SM matter in the bulk for 
$\SU{5}_\mathrm{GUT}$ and $\SO{10}_\mathrm{GUT}$ in the following:

In the $\SU{5}_\mathrm{GUT}$, a complete SM generation is given by the representations $\bar{5}+10+1$ 
of $\SU{5}_\mathrm{GUT}$. What happens if we assume that these matter fields live in the bulk of 
the orbifold? In this case, the singlet (of the right-handed neutrino) is not affected by the 
GUT-breaking boundary condition $P_{\SU{5}}$. For the other $\SU{5}_\mathrm{GUT}$ representations 
each matter field can be a positive or negative eigenstate of $P_{\SU{5}}$, which determines the 
respective zero modes. For example, let us denote the $\bar{5}$ of $\SU{5}$ as $F = (d^c, \ell)$. 
Then, eq.~\eqref{eqn:OrbifoldVEVBC} yields
\begin{subequations}
\begin{eqnarray}
F & \stackrel{P_{\SU{5}}}{\longmapsto} & P_{\SU{5}}\, F ~=~ +F  \quad\Rightarrow\quad F ~\equiv~ \bar{5}^+\to d^c \quad\mathrm{or} \\
F & \stackrel{P_{\SU{5}}}{\longmapsto} & P_{\SU{5}}\, F ~=~ -F  \quad\Rightarrow\quad F ~\equiv~ \bar{5}^-\to \ell\;,
\end{eqnarray}
\end{subequations}
while the $10$ of $\SU{5}_\mathrm{GUT}$ (being an anti-symmetric $5\times 5$ matrix $T$) is subject 
to the boundary conditions $T \mapsto P_{\SU{5}}\,T\,P_{\SU{5}} = \pm T$. In total, we get
\begin{equation}
\bar{5}^+\to d^c,\ \ \bar{5}^-\to \ell, \ \ 10^+\to u^c + e^c,\ \ 10^-\to q, \ \ 1^+\to n\;.
\end{equation}
To have the full SM matter content after compactification, we have to have each eigenstate. 
Consequently, the number of SM matter bulk fields before compactification must be duplicated. 

In the $\SO{10}_\mathrm{GUT}$, a complete SM generation fits into the $16$ of $\SO{10}_\mathrm{GUT}$. 
Depending on the eigenstate of each boundary condition $P_\mathrm{GG}$ and $P_\mathrm{PS}$ we 
obtain the zero modes
\begin{equation}
16^{++}\to d^c + n,\ \ 16^{+-}\to \ell,\ \ 16^{-+}\to q,\ \ 16^{--}\to u^c + e^c\;.
\end{equation}
Hence, if the SM matter is supposed to originate from the extra-dimensional bulk we need four 
copies of $16$-plets in the orbifold bulk for a single generation of SM quarks and leptons. Note 
that split GUT multiplets can be beneficial to explain the different masses for charged leptons and 
down quarks.

Beside GUT-breaking boundary conditions, SM matter fields will in general be subject to the boundary 
conditions of the flavour group $\SU{3}_\mathrm{fl}$. Obviously, $\SU{3}_\mathrm{fl}$ singlets are 
not affected by the flavour-breaking boundary conditions. However, in order to get non-trivial 
predictions from the $\SU{3}_\mathrm{fl}$ flavour group, we assume that some SM matter fields 
transform as $\SU{3}_\mathrm{fl}$ triplets corresponding to the number of three generations. Then, 
in order to keep the structure in the fermion mass matrices dictated by the flavour symmetry, these 
flavour-triplets must be kept as triplets after compactification. Therefore, the 
$\SU{3}_\mathrm{fl}$ matter triplets must be localized at zero-dimensional fixed points in the 
extra dimensions of the orbifold with trivial centralizers such that they are not subject to 
flavour-breaking boundary conditions.

\section[Flavour from a T6/Z2xZ2 orbifold]{\boldmath Flavour from a $\mathbb{T}^6/\Z{2}\times\Z{2}$ orbifold\unboldmath}
\label{sec:Z2xZ2Orbifold}

Next, we consider a ten-dimensional theory with $\mathcal{N}=1$ supersymmetry compactified on 
a $\mathbb{T}^6/\Z{2}\times\Z{2}$ orbifold~\cite{Forste:2004ie,Blaszczyk:2009in} to four-dimensional 
space-time. In this case, the six-dimensional torus $\mathbb{T}^6$ can be chosen to be factorized 
$\mathbb{T}^6 = \mathbb{T}^2 \times \mathbb{T}^2 \times \mathbb{T}^2$, where each two-torus 
$\mathbb{T}^2$ is specified by two vectors
\begin{equation}
e_{2a-1} \quad\mathrm{and}\quad e_{2a} \quad\mathrm{for}\quad a \in\{1,2,3\}\;.
\end{equation}
Then, the point group $P \cong \Z{2}\times\Z{2}$ is generated by 
\begin{equation}
z ~\mapsto~ \vartheta\,z \quad\mathrm{and}\quad z ~\mapsto~ \omega\,z\;,
\end{equation}
where $z=(z^1,z^2,z^3) \in \mathbbm{C}^3$ denotes the complex coordinates on the three two-tori and 
$\vartheta$ and $\omega$ are given by
\begin{equation}
\vartheta ~=~ \begin{pmatrix} 1&0&0\\0&-1&0\\0&0&-1\end{pmatrix} \quad\mathrm{and}\quad  \omega ~=~ \begin{pmatrix} -1&0&0\\0&1&0\\0&0&-1\end{pmatrix}\;.
\end{equation}
Since $\vartheta, \omega \in \SU{3}$, four-dimensional $\mathcal{N}=1$ supersymmetry can be 
preserved in this orbifold.

This orbifold has $16+16+16=48$ inequivalent fixed point sets $\mathrm{F}_g$ corresponding 
to the constructing elements $g$ being
\begin{subequations}\label{eqn:Z2xZ2ConstructingElements}
\begin{eqnarray}
&& \left(\vartheta,       \sum_{a=2,3} \left(n_{2a-1} e_{2a-1} + n_{2a} e_{2a}\right)\right) \quad\mathrm{where}\quad n_3, n_4, n_5, n_6 \in \{0,1\}\;,\\
&& \left(\omega,          \sum_{a=1,3} \left(n_{2a-1} e_{2a-1} + n_{2a} e_{2a}\right)\right) \quad\mathrm{where}\quad n_1, n_2, n_5, n_6 \in \{0,1\}\;,\\
&& \left(\vartheta\omega, \sum_{a=1,2} \left(n_{2a-1} e_{2a-1} + n_{2a} e_{2a}\right)\right) \quad\mathrm{where}\quad n_1, n_2, n_3, n_4 \in \{0,1\}\;.
\end{eqnarray}
\end{subequations}
Note that each fixed point set $\mathrm{F}_g$ is two-dimensional, e.g.\
\begin{subequations}\label{eq:Z2xZ2FixedPointSets}
\begin{eqnarray}
\mathrm{F}_g & = & \{(z^1,0,0) ~|~ z^1 \in\mathbbm{C}\} \qquad\mathrm{for}\qquad g ~=~ (\vartheta, 0)\;,\\
\mathrm{F}_g & = & \{(0,z^2,0) ~|~ z^2 \in\mathbbm{C}\} \qquad\mathrm{for}\qquad g ~=~ (\omega, 0)\;,\\
\mathrm{F}_g & = & \{(0,0,z^3) ~|~ z^3 \in\mathbbm{C}\} \qquad\mathrm{for}\qquad g ~=~ (\vartheta\omega, 0)\;,
\end{eqnarray}
\end{subequations}
see figure~\ref{fig:T6Z2Z2Orbifold} for a three-dimensional illustration.

\begin{figure}[!t!]
\begin{center}
\includegraphics[width=0.8\textwidth]{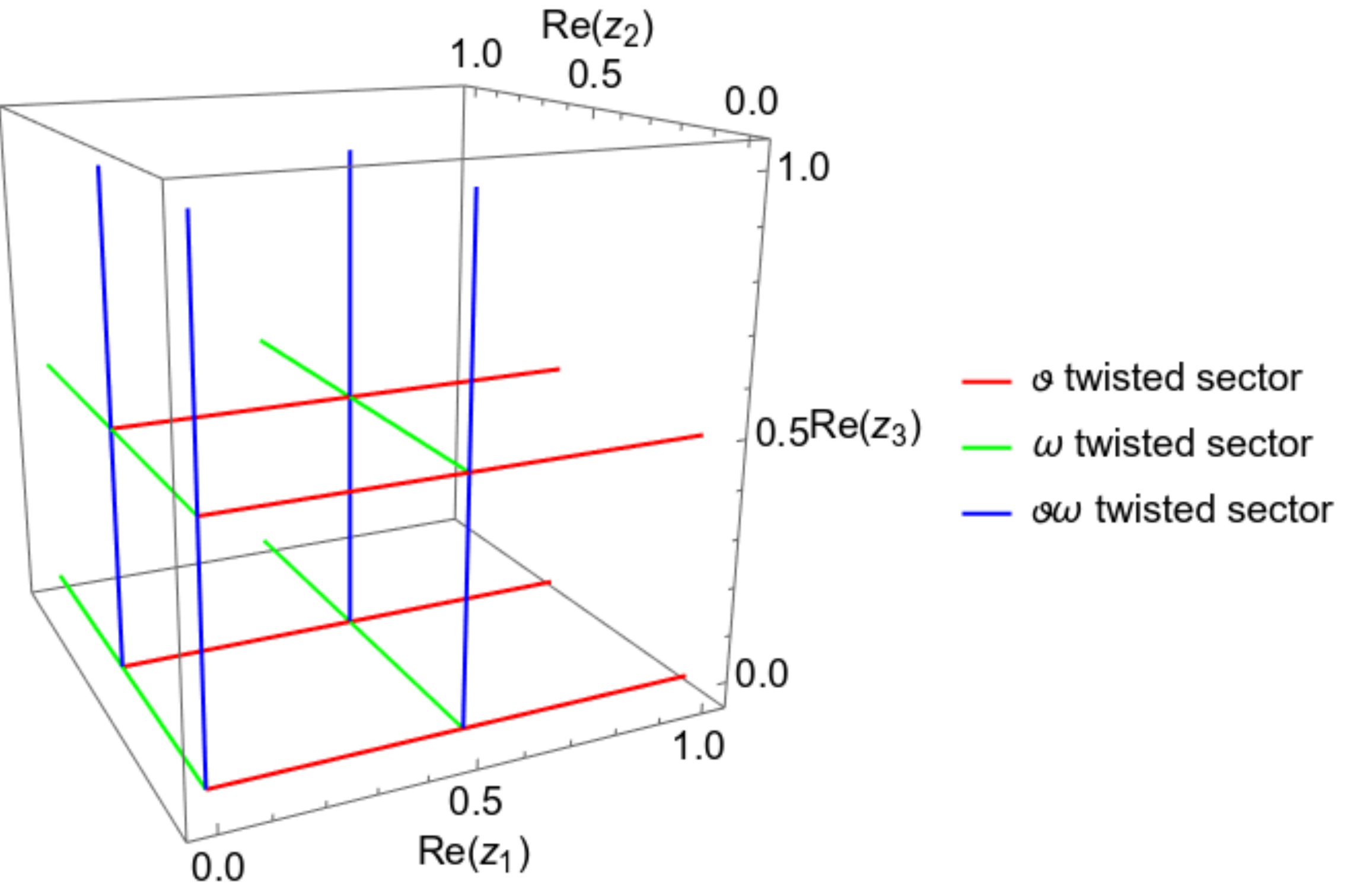}
\caption{Illustration of the six-dimensional $\mathbb{T}^6/\Z{2}\times\Z{2}$ orbifold with an 
orthonormal six-torus, projected onto $(\mathrm{Re}(z_1), \mathrm{Re}(z_2), \mathrm{Re}(z_3))$. The 
orbifold has $16+16+16$ two-dimensional fixed tori corresponding to the $\vartheta$, $\omega$ and 
$\vartheta\omega$ twisted sectors, respectively. They become $4+4+4$ fixed lines in this 
projection, c.f.\ ref.~\cite{Vaudrevange:2010st}.}
\label{fig:T6Z2Z2Orbifold}
\end{center}
\end{figure}

To define a field theory on this $\mathbb{T}^6/\Z{2}\times\Z{2}$ orbifold, we have to choose a 
(gauge) embedding $R_{\vartheta}$, $R_{\omega}$, and $R_{e_i}$ for each generator of the space 
group $S$, i.e.\ for each $(\vartheta, 0)$, $(\omega, 0)$, and $(\Id, e_i)$ for 
$i \in\{1,\ldots,6\}$. We have to ensure that this embedding satisfies the following conditions, 
obtained from eq.~\eqref{eqn:grouphomomorphism}, 
\begin{equation}\label{eqn:Z2xZ2ConditionsA}
\left(R_{\vartheta}\right)^2 ~=~ \Id\;, \quad \left(R_{\omega}\right)^2 ~=~ \Id\;, \quad R_{\vartheta}\, R_{\omega} ~=~  R_{\omega}\,R_{\vartheta}\;,
\end{equation}
and
\begin{subequations}\label{eqn:Z2xZ2ConditionsB}
\begin{align}
R_{e_i}\, R_{e_j}                     & =  R_{e_j}\,R_{e_i}           & \mathrm{for} & \; i,j \in\{1,\ldots,6\}\;, \label{eqn:Z2xZ2Condition4}\\
R_{\vartheta}\,R_{e_i}\,R_{\vartheta} & =  \left(R_{e_i}\right)^{-1}  & \mathrm{for} & \; i   \in\{3,4,5,6\}\;,    \label{eqn:Z2xZ2Condition5}\\
R_{\vartheta}\,R_{e_i}\,R_{\vartheta} & =  R_{e_i}\;,                 & \mathrm{for} & \; i   \in\{1,2\}\;,        \label{eqn:Z2xZ2Condition6}\\
R_{\omega}\,R_{e_i}\,R_{\omega}       & =  \left(R_{e_i}\right)^{-1}  & \mathrm{for} & \; i   \in\{1,2,5,6\}\;,    \label{eqn:Z2xZ2Condition7}\\
R_{\omega}\,R_{e_i}\,R_{\omega}       & =  R_{e_i}\;,                 & \mathrm{for} & \; i   \in\{3,4\}\;,        \label{eqn:Z2xZ2Condition8}
\end{align}
\end{subequations}
see also eqs.~\eqref{eqn:grouphomomorphismConsq1} and~\eqref{eqn:grouphomomorphismConsq2}. Then, 
using the homomorphism property~\eqref{eqn:grouphomomorphism} a general element of the space group 
$g \in S$ has a (gauge) embedding $R_g$ given by
\begin{equation}
g = \left(\vartheta^k\omega^\ell, \sum_{i=1}^6 n_i e_i\right) \;\Leftrightarrow\; R_g = \prod_{i=6}^3 \left(R_{e_i}\right)^{n_i}\, \left(R_{\vartheta}\right)^k \left(R_{\omega}\right)^\ell\;,
\end{equation}
for $k, \ell \in\{0,1\}$.

Compared to the $\mathbb{T}^2/\Z{2}$ orbifold in section~\ref{sec:Z2Orbifold} we have more 
$R$-matrices that could potentially allow us to fix more flavon alignments. However, the 
conditions~\eqref{eqn:Z2xZ2ConditionsA} and~\eqref{eqn:Z2xZ2ConditionsB} are quite restrictive. It 
turns out that if we were to build a flavon-setup as in section~\ref{sec:t2vev} the stringent 
conditions would not allow new useful alignments. We end up with the same alignment capabilities as 
the smaller orbifold $\mathbb{T}^2/\Z{2}$. Furthermore, the $\mathbb{T}^6/\Z{2}\times\Z{2}$ 
orbifold only has two-dimensional fixed tori and no zero-dimensional fixed points. Consequently, 
the centralizers $C_g$ are non-trivial and induce in general projection conditions on localized SM 
matter fields, compare to section~\ref{sec:SMMatterLocalizationZ2}.

\section[Flavour from a T6/S4 orbifold]{\boldmath Flavour from a $\mathbb{T}^6/S_4$ orbifold\unboldmath}
\label{sec:S4Orbifold}

In the previous section, we studied boundary conditions in the $\mathbb{T}^6/\Z{2}\times\Z{2}$ 
orbifold which do not allow for predictive flavour alignments. Hence, we enlarge the orbifolding 
symmetry. We know that the alignment CSD3 can be obtained with the flavour group $S_4$ \cite{King:2016yvg}, and 
noticing that $\mathbb{Z}_2\times\mathbb{Z}_2\subset S_4$, the orbifold $\mathbb{T}^6/S_4$ seems a 
fair choice. 

Consider an orthonormal basis $e_i$ in the six extra dimensions $i\in\{1,\ldots,6\}$, i.e.\ 
\begin{equation}\label{eq:S4Torus}
e_i \cdot e_j = \delta_{ij}\;,
\end{equation}
and define a corresponding orthonormal six-torus $\mathbb{T}^6$ (ignoring the possibility to 
change the overall radius of $\mathbb{T}^6$). In complex coordinates 
$z = (z_1, z_2, z_3) \in \mathbbm{C}^3$ the basis vectors $e_{2a-1}$ and $e_{2a}$ lie in the 
complex plane $z_a$ for $a\in\{1,2,3\}$.

Next, we choose two rotational space group generators $(\vartheta,0)$ and $(\omega,0)$ with the 
following actions on $z = (z_1, z_2, z_3)$
\begin{equation}
\vartheta ~=~ \begin{pmatrix} 0&1&0\\0&0&1\\1&0&0\end{pmatrix} \quad\mathrm{and}\quad  \omega ~=~ \begin{pmatrix} 1&0&0\\0&0&1\\0&-1&0\end{pmatrix}\;,
\end{equation}
see appendix C.2 in ref.~\cite{Fischer:2012qj}. One can check that $\vartheta$ and $\omega$ generate 
the permutation group $S_4$, i.e.
\begin{equation}\label{eq:S4Presentation}
S_4 ~\cong~ \langle \vartheta, \omega ~|~ \omega^4 ~=~ \vartheta^3 ~=~ (\vartheta\, \omega)^2 ~=~ \Id \rangle\;.
\end{equation}
Then, the $\mathbb{T}^6/S_4$ orbifold is defined as the quotient space
\begin{equation}
z ~\sim~ \vartheta\,z \quad\mathrm{and}\quad z ~\sim~ \omega\,z
\end{equation}
of the six-torus~\eqref{eq:S4Torus}. Since $\vartheta, \omega \in \SU{3}$, four-dimensional 
$\mathcal{N}=1$ supersymmetry can be preserved in this orbifold.

According to eq.~\eqref{eq:ConjugacyClasses} in appendix~\ref{app:SpaceGroup} the conjugacy classes 
(of the space group) give rise to the distinct sectors of the theory. Therefore, as a first step one 
needs to determine the conjugacy classes of the point group. As a result, the 24 elements of the 
$S_4$ point group decompose into five conjugacy classes, being
\begin{subequations}\label{eqn:S4ConjugacyClasses}
\begin{eqnarray}
\big[ \Id \big]               & = & \{ \Id \}\;, \\
\big[ \vartheta \big]         & = & \{ \vartheta, \vartheta^2, \vartheta\,\omega^2, \omega\,\vartheta^2\omega, \vartheta\,\omega^2\vartheta, \vartheta^2\omega^2, \omega^2\vartheta^2, \omega^2\vartheta \}\;, \\
\big[ \omega \big]            & = & \{ \omega, \omega\,\vartheta^2, \vartheta^2\omega, \vartheta^2\omega\,\vartheta, \vartheta\,\omega\,\vartheta, \vartheta\,\omega\,\vartheta^2 \}\;, \\
\big[ \omega^2 \big]          & = & \{ \omega^2, \vartheta\,\omega^2\vartheta^2, \vartheta^2\omega^2\vartheta \}\;, \\
\big[ \vartheta\,\omega \big] & = & \{ \vartheta\,\omega, \omega\,\vartheta, \vartheta\,\omega^2\vartheta^2\omega, \vartheta^2\omega\,\vartheta^2, \omega^2\vartheta^2\omega, \omega\,\vartheta^2\omega^2 \}\;.
\end{eqnarray}
\end{subequations}
Subsequently, a full analysis of the conjugacy classes of the space group~\cite{Fischer:2013qza} 
reveals the distinct sectors as given in appendix~\ref{app:S4OrbifoldDetails} and indicated in 
table~\ref{tab:S4Sectors} by the so-called Hodge numbers $(h^{(1,1)},h^{(2,1)})$. In addition, 
figure~\ref{fig:T6S4Orbifold} illustrates the setup. We can interpret the Hodge numbers as follows: 
$h^{(1,1)}$ counts the number of distinct fixed point sets in the various twisted sectors, e.g.\ 
there are ten distinct fixed point sets in the $\omega^2$ twisted sector. As a remark, $h^{(2,1)}$ 
counts how many of the $h^{(1,1)}$ fixed point sets are two-dimensional fixed tori where each 
two-torus is parametrised by a \emph{non-frozen} complex structure modulus (in this case, one can 
modify the angle between the two basis vectors of the two-torus freely). In contrast, a twisted 
sector with $h^{(1,1)} > h^{(2,1)}$ contains $h^{(1,1)} - h^{(2,1)}$ fixed point sets that are 
either zero-dimensional points or two-dimensional tori but with a frozen complex structure.

\begin{table}[!t]
\begin{center}
\begin{tabular}{|c|c|c|c|c|c|}
\hline
twisted sector of              & $(h^{(1,1)},h^{(2,1)})$ & eigenvalues & extra      & centralizer        & generators of \\ 
constr. element $g$            &                         & of twist    & dim.       & $C_g$              & centralizer \\
\hline
\hline
$\big[ \Id \big]$              & $( 1,1)$ & $( 1, 1, 1)$                & 6          & $S_4$              & $\vartheta$, $\omega$\\
\hline
$\big[ \vartheta \big]$        & $( 1,1)$ & $( 1,\alpha^2, \alpha^4)$   & 2          & $\Z{3}$            & $\vartheta$\\
$\big[ \omega \big]$           & $( 4,4)$ & $( 1,\I,-\I)$               & 2          & $\Z{4}$            & $\omega$ \\
$\big[ \omega^2 \big]$         & $(10,0)$ & $( 1,-1,-1)$                & 2          & $D_8$              & $\omega$, $\vartheta^2 \omega\,\vartheta^2$\\
$\big[ \vartheta \omega \big]$ & $( 4,0)$ & $( 1,-1,-1)$                & 2          & $\Z{2}\times\Z{2}$ & $\vartheta \omega$, $\vartheta^2\,\omega^2\,\vartheta$ \\
\hline
\end{tabular}
\caption{Details of the $S_4$ orbifold: $\alpha = \exp(\frac{2\pi\I}{6})$. $D_8$ is the dihedral 
group of order 8. The generators of the centralizer correspond to one specific $g$, for example, in 
the $\omega^2$ twisted sector the centralizer is generated by $\omega$ and $\vartheta^2 \omega\,\vartheta^2$.}
\label{tab:S4Sectors}
\end{center}
\end{table}

\begin{figure}[!t!]
\begin{center}
\includegraphics[width=0.8\textwidth]{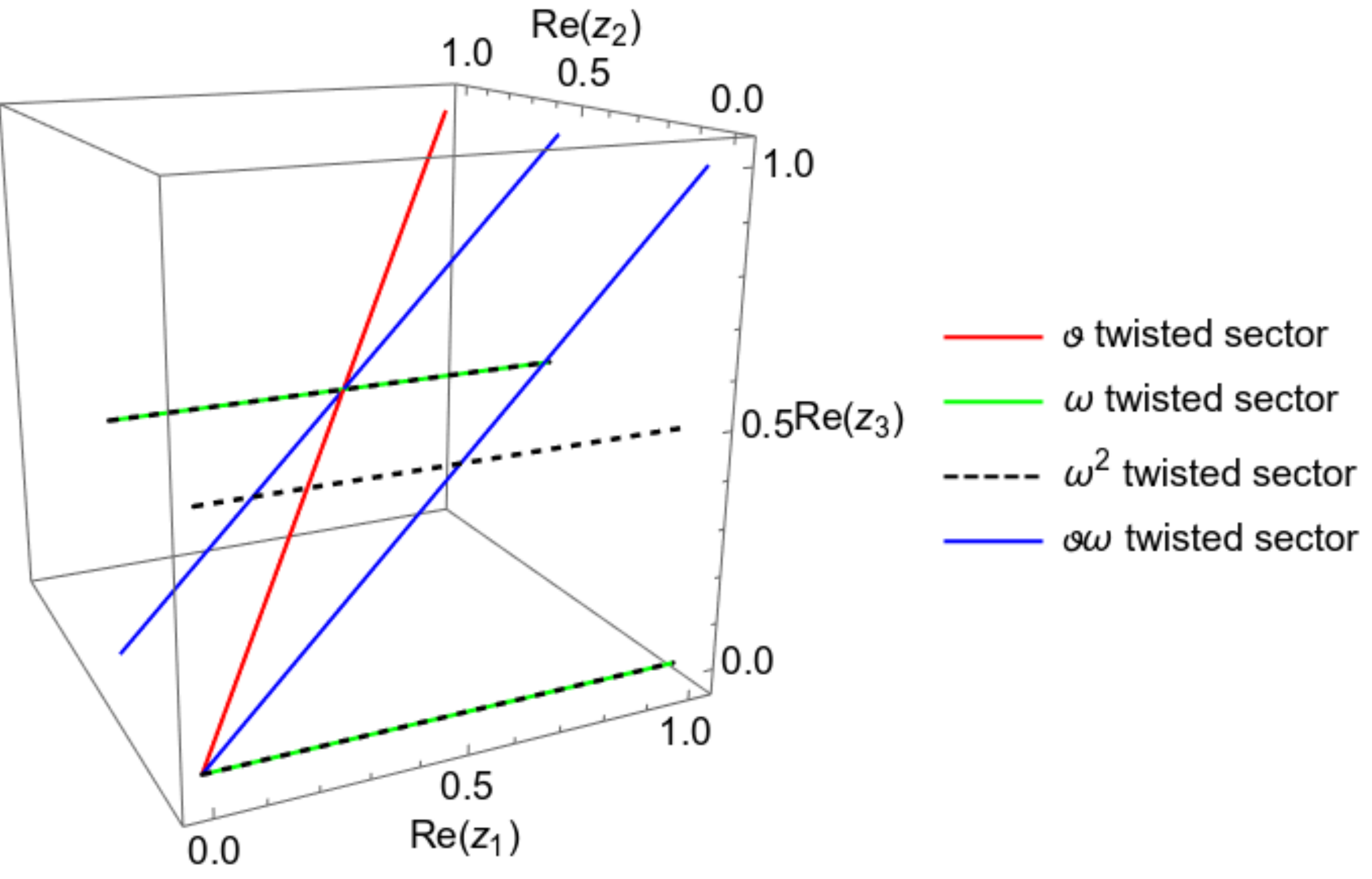}
\caption{Illustration of the six-dimensional $\mathbb{T}^6/S_4$ orbifold projected onto 
$(\mathrm{Re}(z_1), \mathrm{Re}(z_2), \mathrm{Re}(z_3))$. The orbifold has $1+4+10+4$ two-dimensional 
fixed tori corresponding to the $\vartheta$, $\omega$, $\omega^2$ and $\vartheta\omega$ twisted sectors, 
respectively. They become $1+2+3+2$ fixed lines in this projection.}
\label{fig:T6S4Orbifold}
\end{center}
\end{figure}

After fixing the geometry, we have to choose the gauge embeddings $R_\vartheta$, $R_\omega$, and 
$R_{e_i}$ corresponding to the twists $(\vartheta,0)$ and $(\omega,0)$ and the translations 
$(\Id,e_i)$, respectively. The twists $(\vartheta,0)$ and $(\omega,0)$ generate the permutation group 
$S_4$. Since $R$ must be a group homomorphism, see eq.~\eqref{eqn:grouphomomorphism}, $R_\vartheta$ 
and $R_\omega$ can be chosen to generate also $S_4$ or a subgroup thereof (for example, ignoring 
world-sheet modular invariance from string theory one could also choose 
$R_\vartheta = R_\omega = \Id$).

In order to render the gauge embeddings of twists and translations fully compatible, we have to find 
matrices $R_\vartheta$, $R_\omega$, $R_{e_1}$ and $R_{e_2}$ such that
\begin{subequations}\label{eqn:S4Conditions}
\begin{eqnarray}
(R_\omega)^4 ~=~ (R_\vartheta)^3 & = &  (R_{\vartheta\, \omega})^2 ~=~ \Id \;, \qquad\mathrm{i.e.}\  R_\vartheta,\, R_\omega\, \ \mathrm{generate}\ S_4,\\
R_{e_1}\, R_{e_2}               & = & R_{e_2}\,R_{e_1}\;,\\
R_\omega\, R_{e_1}              & = & R_{e_1}\,R_\omega\;,\\
R_\omega\, R_{e_2}              & = & R_{e_2}\,R_\omega\;,\\
\left(R_{\vartheta^2\, \omega\, \vartheta^2}\right)\, R_{e_1} & = & R_{e_1}^{-1}\,\left(R_{\vartheta^2\, \omega\, \vartheta^2}\right) \quad\Leftrightarrow\quad \left(\left(R_{\vartheta^2\, \omega\, \vartheta^2}\right)\, R_{e_1}\right)^2 = \Id\;,\\
\left(R_{\vartheta^2\, \omega\, \vartheta^2}\right)\, R_{e_2} & = & R_{e_2}^{-1}\,\left(R_{\vartheta^2\, \omega\, \vartheta^2}\right) \quad\Leftrightarrow\quad \left(\left(R_{\vartheta^2\, \omega\, \vartheta^2}\right)\, R_{e_2}\right)^2 = \Id\;,
\end{eqnarray}
\end{subequations}
where the matrix $R_{\vartheta^2\, \omega\, \vartheta^2} = (R_\vartheta)^2\, R_\omega\, (R_\vartheta)^2$ is of 
order 2. If these conditions are satisfied we can define $R_{e_i}$ for $i=3,4,5,6$ using 
eqs.~\eqref{eq:S4DefE3E4E5E6} in appendix~\ref{app:S4OrbifoldDetails}. This choice will satisfy all 
conditions, i.e.\ those from the presentation of $S_4$ in eq.~\eqref{eq:S4Presentation} and, 
additionally, those from eqs.~\eqref{eq:S4ConditionThetaE} and~\eqref{eq:S4ConditionOmegaE}.

\subsection{\boldmath Alignment of flavon VEVs\unboldmath}

We assume to have an $\SU{3}_\mathrm{fl}$ gauged flavour symmetry and choose the gauge embedding of 
the $S_4$ transformations using the known generators~\cite{King:2011zj} 
\begin{equation}
S ~=~\frac{1}{3}\begin{pmatrix} -1 & 2&2 \\ 2&-1&2 \\2&2&-1 \end{pmatrix}\; ,\ \ T~=~\begin{pmatrix} 1&0&0\\ 0&\alpha^4 &0 \\ 0&0&\alpha^2 \end{pmatrix}\;,\ \ 
U ~=~ \begin{pmatrix} 1&0&0\\0&0& 1 \\ 0&1&0 
\end{pmatrix}\;,
\label{eq:a4gen}
\end{equation} 
as
\begin{equation}\label{eq:S4ChoiceForGaugeEmbedding}
R_\vartheta ~=~T \quad,\quad R_\omega ~=~ UTS ~=~ \frac{1}{3}\begin{pmatrix} -1 & 2&2\\2\alpha^2&2\alpha^2&-\alpha^2 \\ 2\alpha^4&-\alpha^4&2\alpha^4  \end{pmatrix} \quad\mathrm{and}\quad R_{e_i} ~=~ \Id\;,
\end{equation}
where $\alpha =\exp(\frac{2\pi\I}{6})$. This choice satisfies all gauge embedding conditions for an $S_4$ orbifold.

Next, we choose to have three flavons $\phi_1$, $\phi_2$, and $\phi_3$ (each being a triplet of the 
$\SU{3}_\mathrm{fl}$ flavour group) and localize them in different sectors of the 
$\mathbb{T}^6/S_4$ orbifold as listed in table~\ref{tab:LocalizationOfFlavons}. Let us begin with 
specifying the flavon $\phi_1$ in great detail so that our discussion for $\phi_2$ and $\phi_3$ can 
be shorter later on. We want the flavon $\phi_1$ to be subject to the boundary condition 
$R_\vartheta\,R_\omega$. Hence, we localize it in the $\vartheta\,\omega$ twisted sector (e.g.\ on 
the fixed torus $z_\mathrm{f}\in\mathbbm{C}^3$ given by the solutions of 
$\vartheta\,\omega\,z_\mathrm{f}=z_\mathrm{f}$). In order to identify all boundary conditions that 
act on $\phi_1$ we have to compute the centralizer of $\vartheta\,\omega$, i.e.\ we have to 
identify all elements of $S_4$ that commute with $\vartheta\,\omega$. It turns out that the 
centralizer of $\vartheta\,\omega$ is generated by 
\begin{equation}
\vartheta\,\omega \quad\mathrm{and}\quad \vartheta^2\,\omega^2\,\vartheta\;,
\end{equation}
and corresponds to $\Z{2}\times\Z{2}$. Consequently, the flavon $\phi_1$ will feel the boundary 
conditions $R_g$ of all elements $g$ of the centralizer (up to 
some phases as introduced in section~\ref{sec:VEValignment} that can be chosen freely). Hence, 
$\phi_1$ is subject to
\begin{equation}
R_\vartheta\,R_\omega\,\langle\phi_1\rangle ~=~ \pm\langle\phi_1\rangle \quad\mathrm{and}\quad \left(R_\vartheta\right)^2\,\left(R_\omega\right)^2\,R_\vartheta\,\langle\phi_1\rangle ~=~ \pm\langle\phi_1\rangle\;,
\end{equation}
where the $\pm$ signs in both conditions can be chosen independently.

If this does not work out, there are ways to make the centralizer smaller. But then $S_4$ orbifold 
becomes more complicated. For example, one can use various different six-tori that can not be written 
as $\mathbb{T}^2\times \mathbb{T}^2 \times \mathbb{T}^2$, c.f.\ ref.~\cite{Fischer:2012qj}.

\begin{table}[!t]
\begin{center}
\begin{tabular}{|c|c|c|c|}
\hline
flavon   & localization                 & centralizer        & generators of centralizer \\ 
\hline
\hline
$\phi_1$ & $\vartheta\,\omega             ~\in~ [\vartheta\,\omega ]$ & $\Z{2}\times\Z{2}$ & $\vartheta\,\omega$ and $\vartheta^2\,\omega^2\,\vartheta$\\
$\phi_2$ & $\omega\,\vartheta^2\,\omega^2 ~\in~ [\vartheta\,\omega ]$ & $\Z{2}\times\Z{2}$ & $\omega\,\vartheta^2\,\omega^2$ and $\vartheta^2\,\omega^2\,\vartheta$\\
$\phi_3$ & $\vartheta                     ~\in~ [\vartheta ]$         & $\Z{3}$            & $\vartheta$\\
\hline
\end{tabular}
\caption{Localizations of the three flavons $\phi_1$, $\phi_2$ and $\phi_3$ in the various sectors of the $\mathbb{T}^6/S_4$ orbifold and their centralizers, which indicates which boundary condition the respective flavon is feeling.}
\label{tab:LocalizationOfFlavons}
\end{center}
\end{table}

After we have chosen the localization of each flavon, the flavon VEVs must comply with the 
respective boundary conditions. We assume that the flavons obtain a non-vanishing VEV through some 
other mechanism. However, the alignment of the flavon VEV in flavour space is fixed to a specific 
direction through the boundary conditions.

In more detail, the flavon $\phi_3$ is chosen to be localized in the $\vartheta$ sector. Its VEV 
must comply with the boundary condition $R_\vartheta=T$. It has the freedom of having any of the 
three phases $\alpha^{2n}$ for $n=0,1,2$, and we choose it to be $\alpha^{4}$ so that
\begin{equation}
\langle\phi_3\rangle=\alpha^4 R_\vartheta\langle\phi_3\rangle=\alpha^4 T\langle\phi_3\rangle=\begin{pmatrix} \alpha^4&0&0\\ 0&\alpha^2 &0 \\ 0&0& 1\end{pmatrix}\braket{\phi_3}\ \ \rightarrow\ \ \langle\phi_3\rangle\propto \left(\begin{array}{c}0 \\ 0 \\ 1
\end{array}\right),
\end{equation}
which aligns the VEV completely. This VEV must be invariant (up to a phase) under the 
full centralizer of $\vartheta$, which is generated by $\vartheta$ itself, so it is consistent.

The flavon $\phi_2$ is chosen to be localized in the $\omega\vartheta^2\omega^2$ sector, so that 
its VEV must be invariant under $R_\omega R_\vartheta^2 R_\omega^2=U$, up to a sign, which we 
choose to be negative. This enforces the VEV to be
\begin{equation}\label{eq:S4FlavonPhi2}
\langle\phi_2\rangle=-R_\omega R_\vartheta^2 R_\omega^2\langle\phi_2\rangle=-U\langle\phi_2\rangle=\begin{pmatrix} -1&0&0\\ 0&0 &-1 \\ 0&-1& 0\end{pmatrix}\braket{\phi_2}\ \ \rightarrow\ \ \langle\phi_2\rangle\propto \left(\begin{array}{c}0 \\ 1 \\ -1
\end{array}\right),
\end{equation}
which aligns the VEV completely. The VEV must also be invariant (up to a sign) with the 
corresponding centralizer, which in this case is generated by $\omega\,\vartheta^2\,\omega^2$ and 
$\vartheta^2\,\omega^2\,\vartheta$. Hence, the VEV eq.~\eqref{eq:S4FlavonPhi2} must also be 
invariant under the boundary condition using $R_\vartheta^2 R_\omega^2 R_\vartheta=S,$ up to a 
sign. We choose the sign to be negative (the positive sign would force the VEV to vanish) so that 
\begin{equation}
\langle\phi_2\rangle=-R_\vartheta^2 R_\omega^2 R_\vartheta\langle\phi_2\rangle=-S\langle\phi_2\rangle=\frac{-1}{3}\begin{pmatrix} -1 & 2&2 \\ 2&-1&2 \\2&2&-1 \end{pmatrix}\braket{\phi_2}\ \ \rightarrow\ \ \langle\phi_2\rangle\propto \left(\begin{array}{c}a \\ b \\ -a-b
\end{array}\right),
\end{equation}
with arbitrary $a,b$. This alignment is compatible with the previous condition when $a=0$. This fixes the VEV $\phi_2$ completely and consistently through boundary conditions.

The flavon $\phi_1$ obtains a VEV that, due to the choice of the localization in the sector $\vartheta\omega$, must be invariant under the boundary conditions $R_\vartheta R_\omega=SU$ up to a sign. We choose the positive sign so that
\begin{equation}
\langle\phi_1\rangle=R_\vartheta R_\omega\langle\phi_1\rangle=SU\langle\phi_1\rangle=\frac{1}{3}\begin{pmatrix} -1 & 2&2 \\2&2&-1 \\ 2&-1&2 \end{pmatrix}\braket{\phi_1}\ \ \rightarrow\ \ \langle\phi_1\rangle\propto \left(\begin{array}{c}a \\ b \\ 2a-b
\end{array}\right).
\end{equation}
This VEV is aligned in the general CSDn direction which is defined with $a=1,\ b=n$.
It must also comply with the boundary conditions of the centralizer, up to a sign, which is generated by $\vartheta\,\omega$ and $\vartheta^2\,\omega^2\,\vartheta$.  This VEV must also be invariant under the boundary condition $R_\vartheta^2 R_\omega^2 R_\vartheta=S,$ up to a sign. We choose the sign to be positive (the negative sign would force the VEV to vanish) so that 
\begin{equation}
\langle\phi_1\rangle=R_\vartheta^2 R_\omega^2 R_\vartheta\langle\phi_1\rangle=S\langle\phi_1\rangle=\frac{1}{3}\begin{pmatrix} -1 & 2&2 \\ 2&-1&2 \\2&2&-1 \end{pmatrix}\braket{\phi_1}\ \ \rightarrow\ \ \langle\phi_1\rangle\propto \left(\begin{array}{c}1 \\ 1 \\ 1
\end{array}\right),
\end{equation}
which is consistent with the previous condition fixing $a=b=1$. This is the CSD1 alignment which is 
widely used in the tribimaximal (TBM) alignment~\cite{King:2005bj}.

We conclude that the flavon VEV alignments can be fixed completely and consistently to the TBM 
alignment and we can arrange for a situation with three flavons $\phi_i$ such that
\begin{equation}\label{eq:S4VEVAlignments}
\langle\phi_1\rangle\propto \left(\begin{array}{c}1 \\ 1 \\ 1
\end{array}\right),\ \ \langle\phi_2\rangle\propto \left(\begin{array}{c}0 \\ 1 \\ -1
\end{array}\right),\ \ \langle\phi_3\rangle\propto \left(\begin{array}{c}0 \\ 0 \\ 1
\end{array}\right).
\end{equation}
These flavons are enough to fit all masses predictibly, specially in the lepton sector.

\subsection{Roto-translations}

Another realization of the $S_4$ orbifold is based on a space group with roto-translations
\begin{equation}
\left(\vartheta, \frac{1}{4}(e_1+e_3)\right) \qquad\mathrm{and}\qquad \left(\omega, \frac{1}{4}(e_1+3e_2)\right)\;.
\end{equation}
In this case, there are only three sectors corresponding to $[\Id]$, $[\vartheta]$ and $[\vartheta\,\omega]$, 
where the later two, twisted sectors have trivial centralizers. Thus, one can localize the flavons 
$\phi_1$ and $\phi_2$ in the sector $[\vartheta\,\omega]$, while $\phi_3$ is localized in $[\vartheta]$.

This way we can obtain the general CSDn complete alignment. However, we did not find a mechanism 
to fix $n=2,3$ so that we have a highly predictive fermion mass setup as in 
ref.~\cite{Bjorkeroth:2017ybg,Bjorkeroth:2015ora,deAnda:2018oik}.

\subsection{GUT breaking}
Up to now, we have only chosen specific embedding matrices $R_{\vartheta}$ and $R_{\omega}$ in 
eq.~\eqref{eq:S4ChoiceForGaugeEmbedding} to align flavons in the CSDn or TBM alignment. We have not 
fixed any $R_{e_i}$ in the process. From eqs.~\ref{eqn:S4Conditions}, we see that there are only 
two free matrices ($R_{e_{1}}$ and $R_{e_{2}}$) to choose and they must comply with their specific 
conditions. Hence, one option is to choose $R_{e_{1}}$ and $R_{e_{2}}$ to break the GUT but not 
flavour. 

In this case, the matrices $R_{e_{1}}$ and $R_{e_{2}}$ would commute with $R_{\vartheta}$ and 
$R_{\omega}$. The only remaining constraints from eqs.~\ref{eqn:S4Conditions} state that 
$R_{e_{1}}$ and $R_{e_{2}}$ must commute with each other and be of order two. Hence, they 
generate $\Z{2}\times\Z{2}$. Consequently, we are allowed to apply the GUT breaking mechanism as 
discussed in section~\ref{sec:GUTZ2}.

\subsection{SM fermion localization}

We have shown that this $S_4$ orbifold is enough to align three different flavons in the TBM or the 
CSDn setups and break GUTs.

However, we want some of the SM fermions (e.g.\ the charged leptons $\ell$) to form a triplet 
of $\SU{3}_\mathrm{fl}$ flavour. Thus, these fermions must be located where they are not affected 
by the flavour breaking conditions associated to $R_{\vartheta}$ or $R_{\omega}$. In this $S_4$ 
orbifold the only places to locate fields are the bulk and invariant tori, see 
table~\ref{tab:S4Sectors}, which all are affected by the conditions $R_{\vartheta}$ or 
$R_{\omega}$. Hence, some components of $\SU{3}_\mathrm{fl}$ flavour triplets for SM matter are 
necessarily projected out in this $S_4$ orbifold. This would destroy any predictability coming from 
the flavon alignments~\eqref{eq:S4VEVAlignments}. Consequently, we move on to another orbifold that 
allows for suitable fermion localizations.

\section[Flavour from a T6/Delta(54) orbifold]{\boldmath Flavour from a $\mathbb{T}^6/\Delta(54)$ orbifold\unboldmath}
\label{sec:Delta54}

We want to enlarge the orbifolding symmetry once again, to allow a place to locate the fermions consistently but keep the alignments we have achieved. 
We can note that $S_4\simeq \Delta(24)$  so that we can continue in the $\Delta(6n^2)$ discrete group series by choosing the next one,
$\Delta(54)$.

Consider a factorized six-torus $\mathbb{T}^6 = \mathbb{T}^2 \times\mathbb{T}^2 \times\mathbb{T}^2$ 
where the $a$-th torus is spanned by basis vectors $e_{2a-1}$ and $e_{2a}$ of length $r$, where 
\begin{equation}
|e_{2a-1}| ~=~ |e_{2a}| ~=~ r \quad\mathrm{and}\quad e_{2a-1} \cdot e_{2a} = -\frac{r^2}{2} \quad\mathrm{for}\ a = 1,2,3\;.
\end{equation}
In complex coordinates $z = (z_1, z_2, z_3) \in \mathbbm{C}^3$ the basis vectors $e_{2a-1}$ and 
$e_{2a}$ lie in the complex plane $z_a$ for $a=1,2,3$. As a remark, this orbifold has only a single 
K{\"a}hler modulus $T$ which parameterizes the overall size $r$ and the overall 
$B$-field~\cite{Fischer:2013qza}.

Next, we choose three space group generators $(\vartheta,0)$, $(\omega,0)$ and $(\rho,0)$ with the 
following actions on $z = (z_1, z_2, z_3)$
\begin{equation}
\vartheta ~=~ \begin{pmatrix} -1&0&0\\0&0&-1\\0&-1&0\end{pmatrix}, \quad\omega ~=~ \begin{pmatrix} 0&1&0\\0&0&1\\1&0&0\end{pmatrix} \quad\mathrm{and}\quad  \rho ~=~ \begin{pmatrix} 0&\alpha^2&0\\0&0&\alpha^4\\1&0&0\end{pmatrix}\;,
\end{equation}
where $\alpha = \exp(\frac{2\pi\I}{6})$. Since 
\begin{equation}
\vartheta^2 ~=~ \omega^3 ~=~ \left(\omega\vartheta\right)^2 ~=~ \Id\;,
\end{equation}
we see that $\vartheta$ and $\omega$ generate an $S_3$ subgroup. Furthermore, 
\begin{equation}
\omega^2\,\rho  ~=~ \begin{pmatrix} 1&0&0\\0&\alpha^2&0\\0&0&\alpha^4\end{pmatrix} \quad\mathrm{and}\quad  \rho\,\omega^2  ~=~ \begin{pmatrix} \alpha^2&0&0\\0&\alpha^4&0\\0&0&1\end{pmatrix}\;,
\end{equation}
generate a $\Z{3}\times\Z{3}$ subgroup such that, finally, we can write 
$\Delta(54) = \left(\Z{3}\times\Z{3}\right) \rtimes S_3$, see e.g.~\cite{Ishimori:2010au}. Finally, 
since $\vartheta, \omega, \rho \in \SU{3}$, four-dimensional $\mathcal{N}=1$ supersymmetry can be 
preserved in this orbifold.

The 54 elements of $\Delta(54)$ decompose into ten conjugacy classes being
\begin{subequations}\label{eqn:Delta54ConjugacyClasses}
\begin{eqnarray}
\big[ \Id \big]                                    & = & \{ \Id \}\;, \\
\big[ \vartheta \big]                              & = & \{ \vartheta, \vartheta \omega,\rho \vartheta \rho^2,\omega \vartheta,\omega \rho \vartheta \omega,\vartheta \rho \omega,\rho \omega \vartheta,\vartheta \rho^2 \omega,\omega \vartheta \rho \}\;, \\
\big[ \omega \big]                                 & = & \{ \omega,\omega^2,\rho \omega \rho^2,\rho \vartheta \omega \rho \vartheta,\vartheta \omega \rho \vartheta \rho,\rho^2 \omega \rho \}\;, \\
\big[ \rho \big]                                   & = & \{ \rho,\vartheta \rho \vartheta,\omega \rho \omega^2,\vartheta \omega \rho \vartheta \omega,\rho^2,\vartheta \rho^2 \vartheta \}\;, \\
\big[ \vartheta\,\rho \big]                        & = & \{ \vartheta \rho,\rho \vartheta,\rho^2 \vartheta \omega,\vartheta \rho^2 \omega^2,\vartheta \omega \rho^2,\omega \rho \vartheta \omega \rho,\vartheta \omega \rho \omega,\rho \vartheta \omega \rho,\rho \omega \rho^2 \vartheta \}\;, \\
\big[ \omega\,\rho \big]                           & = & \{ \omega \rho,\vartheta \omega \rho \vartheta,\vartheta \rho^2 \vartheta \omega,\rho \omega,\rho^2 \omega^2,\vartheta \rho \omega \vartheta \}\;, \\
\big[ \vartheta\,\omega\,\rho \big]                & = & \{ \vartheta \omega \rho,\omega \rho \vartheta,\vartheta \rho \omega^2,\rho \vartheta \omega,\vartheta \rho^2,\omega \rho \vartheta \rho,\rho^2 \vartheta,\rho \vartheta \rho \omega,\rho \vartheta \rho \}\;, \\
\big[ \omega^2\,\rho \big]                         & = & \{ \omega^2 \rho, \rho^2 \omega,\rho \omega^2,\rho \omega \rho, \omega \rho^2,\omega \rho \omega \}\;, \\
\big[ \left(\vartheta\,\rho\right)^2 \big]         & = & \{ \left(\vartheta\,\rho\right)^2 \}\;, \\
\big[ \left(\vartheta\,\omega\,\rho\right)^2 \big] & = & \{ \left(\vartheta\,\omega\,\rho\right)^2 \}\;.
\end{eqnarray}
\end{subequations}

\begin{table}[!t]
\begin{center}
\begin{tabular}{|c|c|c|c|c|c|}
\hline
twisted sector of                                  & $(h^{(1,1)},h^{(2,1)})$ & eigenvalues       & extra      & centralizer        & generators of \\ 
constr. element $g$                                &                         & of twist          & dim.       & $C_g$              & centralizer \\
\hline
\hline
$\big[ \Id \big]$                                  & $(1,0)$ & $( 1, 1, 1)$                      & 6          & $\Delta(54)$       & $\vartheta$, $\omega$, $\rho$\\
\hline
$\big[ \vartheta \big]$                            & $(2,1)$ &  $(-1,-1, 1)$                     & 2          & $\Z{6}$            & $\rho\vartheta\rho$\\
$\big[ \omega \big]$                               & $(9,0)$ &  $(\alpha^2, \alpha^4, 1)$        & 2          & $\Z{3}\times\Z{3}$ & $\omega$, $\left(\vartheta \rho\right)^2$ \\
$\big[ \rho \big]$                                 & $(1,0)$ &  $(\alpha^2, \alpha^4, 1)$        & 2          & $\Z{3}\times\Z{3}$ & $\rho$, $\vartheta \rho \vartheta$\\
$\big[ \omega \rho \big]$                          & $(1,0)$ &  $(\alpha^2, \alpha^4, 1)$        & 2          & $\Z{3}\times\Z{3}$ & $\omega \rho$, $\left(\vartheta \rho\right)^2$\\
$\big[ \omega^2 \rho \big]$                        & $(1,0)$ &  $(\alpha^2, \alpha^4, 1)$        & 2          & $\Z{3}\times\Z{3}$ & $\omega^2 \rho$, $\left(\vartheta \rho\right)^2$\\
\hline
$\big[ \left(\vartheta \omega \rho\right)^2 \big]$ & $(7,0)$ &  $(\alpha^4, \alpha^4, \alpha^4)$ & 0          & $\Delta(54)$       & $\vartheta$, $\omega$, $\rho$\\
$\big[ \left(\vartheta \rho\right)^2 \big]$        & -       &  $(\alpha^2, \alpha^2, \alpha^2)$ & 0          & $\Delta(54)$       & $\vartheta$, $\omega$, $\rho$\\
$\big[ \vartheta \rho \big]$                       & $(3,0)$ &  $(\alpha^4, \alpha,   \alpha)$   & 0          & $\Z{6}$            & $\vartheta \rho$ \\
$\big[ \vartheta \omega \rho \big]$                & -       &  $(\alpha^2, \alpha^5, \alpha^5)$ & 0          & $\Z{6}$            & $\vartheta \omega \rho$ \\
\hline
\end{tabular}
\caption{The various (twisted) sectors of the $\Delta(54)$ orbifold are labelled by their 
constructing elements $g$. The Hodge numbers $(h^{(1,1)},h^{(2,1)})$ count the number of fixed point 
sets $\mathrm{F}_g$ (and their deformations). Each eigenvalue of $+1$ indicates two extra dimensions 
of $\mathrm{F}_g$ such that, for example, $\mathrm{F}_{\vartheta \rho}$ yields zero-dimensional 
fixed points while $\mathrm{F}_{\vartheta}$ gives two-dimensional fixed tori 
(where $\alpha = \exp(\frac{2\pi\I}{6})$).}
\label{tab:Delta54Sectors}
\end{center}
\end{table}

To define a field theory on this $\Delta(54)$ orbifold, we have to choose 
a (gauge) embedding 
\begin{equation}\label{eqn:RGeneratorsDelta54}
R_{\vartheta}\;, R_{\omega}\;, R_{\rho}\;, \quad\mathrm{and}\quad R_{e_i}\;, \quad\mathrm{for}\quad i ~\in~ \{1,\ldots,6\}\;,
\end{equation}
for each generator of the $\Delta(54)$ space group $S$, i.e.\ for each 
\begin{equation}
(\vartheta, 0)\;, (\omega, 0)\;, (\rho, 0)\;, (\Id, e_i) \quad\mathrm{for}\quad i ~\in~ \{1,\ldots,6\}\;.
\end{equation}
We have to ensure that this embedding satisfies the following conditions, obtained from 
eq.~\eqref{eqn:grouphomomorphism}, 
\begin{subequations}\label{eqn:Delta54ConditionsA}
\begin{eqnarray}
\left(R_{\vartheta}\right)^2 ~=~ \left(R_{\omega}\right)^3 ~=~ \left(R_{\rho}\right)^3 & = & \Id\;\\
R_{\vartheta}\,R_{\omega}\left(R_{\vartheta}\right)^{-1}\left(R_{\omega}\right)^{-1}   & = & R_{\omega}\;,\\
R_{\vartheta}\,R_{\rho}  \left(R_{\vartheta}\right)^{-1}\left(R_{\rho}\right)^{-1}     & = & R_{\omega}\, R_{\rho}\left(R_{\omega}\right)^2\;,\\
R_{\omega}\,   R_{\rho}  \left(R_{\omega}\right)^{-1}   \left(R_{\rho}\right)^{-1}     & = & \left(R_{\vartheta}\, R_{\rho}\right)^2\;,
\end{eqnarray}
\end{subequations}
such that $R_{\vartheta}$, $R_{\omega}$ and $R_{\rho}$ generate $\Delta(54)$ or a subgroup thereof. 
Furthermore, 
\begin{subequations}\label{eqn:Delta54ConditionsB}
\begin{align}
R_{e_i}\, R_{e_j}                                & =  R_{e_j}\,R_{e_i} \;, \label{eqn:Delta54Condition1}\\
R_{\vartheta}\,R_{e_i}\,R_{\vartheta}            & =  R_{\vartheta\,e_i}\;,\label{eqn:Delta54Condition2}\\
R_{\omega}\,R_{e_i}\,\left(R_{\omega}\right)^{-1}& =  R_{\omega\,e_i}\;,   \label{eqn:Delta54Condition3}\\
R_{\rho}\,R_{e_i}\,\left(R_{\rho}\right)^{-1}    & =  R_{\rho\,e_i}\;,     \label{eqn:Delta54Condition4}
\end{align}
\end{subequations}
where $i,j\in\{1,\ldots,6\}$, see also eqs.~\eqref{eqn:grouphomomorphismConsq1}
and~\eqref{eqn:grouphomomorphismConsq2}. Explicitly, 
eqs.~\eqref{eqn:Delta54Condition2},~\eqref{eqn:Delta54Condition3} and~\eqref{eqn:Delta54Condition4} 
read
\begin{subequations}\label{eqn:Delta54ConditionsC1}
\begin{align}
R_{\vartheta}\,R_{e_i}\,R_{\vartheta}             & =  \left(R_{e_{i}}\right)^{-1}   &\mathrm{for}\quad i \in \{1,2\}\;,\\
R_{\vartheta}\,R_{e_i}\,R_{\vartheta}             & =  \left(R_{e_{i+2}}\right)^{-1} &\mathrm{for}\quad i \in \{3,4\}\;,\\
R_{\vartheta}\,R_{e_i}\,R_{\vartheta}             & =  \left(R_{e_{i-2}}\right)^{-1} &\mathrm{for}\quad i \in \{5,6\}\;,\\
\end{align}
\end{subequations}
and
\begin{subequations}\label{eqn:Delta54ConditionsC2}
\begin{align}
R_{\omega}\,R_{e_i}\,\left(R_{\omega}\right)^{-1} & =  R_{e_{i+4}}                   &\mathrm{for}\quad i \in \{1,2\}\;,\\
R_{\omega}\,R_{e_i}\,\left(R_{\omega}\right)^{-1} & =  R_{e_{i-2}}                   &\mathrm{for}\quad i \in \{3,4\}\;,\\
R_{\omega}\,R_{e_i}\,\left(R_{\omega}\right)^{-1} & =  R_{e_{i-2}}                   &\mathrm{for}\quad i \in \{5,6\}\;,\\
\end{align}
\end{subequations}
and
\begin{subequations}\label{eqn:Delta54ConditionsC3}
\begin{align}
R_{\omega^2\,\rho}\,R_{e_i}\,\left(R_{\omega^2\,\rho}\right)^{-1} & =  R_{e_{i}}     &\mathrm{for}\quad i \in \{1,2\}\;,\\
R_{\omega^2\,\rho}\,R_{e_3}\,\left(R_{\omega^2\,\rho}\right)^{-1} & =  R_{e_{4}}     &\;\\
R_{\omega^2\,\rho}\,R_{e_4}\,\left(R_{\omega^2\,\rho}\right)^{-1} & =  \left(R_{e_{3}}\right)^{-1}\,\left(R_{e_{4}}\right)^{-1}   &\;\\
R_{\omega^2\,\rho}\,R_{e_5}\,\left(R_{\omega^2\,\rho}\right)^{-1} & =  \left(R_{e_{5}}\right)^{-1}\,\left(R_{e_{6}}\right)^{-1}   &\;\\
R_{\omega^2\,\rho}\,R_{e_6}\,\left(R_{\omega^2\,\rho}\right)^{-1} & =  R_{e_{5}}\;,  &
\end{align}
\end{subequations}
where we use $R_{\omega^2\,\rho}$ instead of $R_\rho$ in order to keep the 
conditions~\eqref{eqn:Delta54ConditionsC3} simple.

One possibility to solve eqs.~\eqref{eqn:Delta54ConditionsC1},~\eqref{eqn:Delta54ConditionsC2} 
and~\eqref{eqn:Delta54ConditionsC3} is given by assuming that
\begin{equation}
R_\theta\, R_{e_i} ~=~ R_{e_i}\,R_\theta \quad\mathrm{for}\quad i ~\in~ \{1,\ldots,6\}\;.
\end{equation}
for all point group elements $\theta \in \Delta(54)$. Importantly, one can show that in this case 
the (gauge) embeddings of the translations have to be trivial, i.e.
\begin{equation}\label{eq:Delta54eiTrivial}
R_{e_i} ~=~ \Id \quad\mathrm{for}\quad i ~\in~ \{1,\ldots,6\}\;.
\end{equation}
Then, we are left with $R_{\vartheta}$, $R_{\omega}$ and $R_{\rho}$ that have to satisfy 
eq.~\eqref{eqn:Delta54ConditionsA}. In the following, we will choose standard embedding 
$R_{\vartheta} = \vartheta$, $R_{\omega} = \omega$ and $R_{\rho} = \rho$ with gauged flavour 
symmetry $\mathcal{G}=\SU{3}_\mathrm{fl}$, c.f.\ section~\ref{sec:StandardEmbedding}.

\subsection{VEV alignment}

Consider a (flavon) field $\Phi_g(x,z)$ localized at $z\in\mathrm{F}_g$ with constructing element 
$g \in S$. We denote the order of $g$ by $N_g$, i.e.\ $g^{(N_g)}=\Id$. Then, the field $\Phi_g(x,z)$ 
has to satisfy the boundary conditions
\begin{subequations}\label{eqn:Delta54BC}
\begin{eqnarray}
\Phi_g(x,g\,z) & = & \exp\left(\frac{2\pi\I\, k}{N_g}\right)\, R_g\,\Phi_g(x,z)\;, \\
\Phi_g(x,h\,z) & = & \exp\left(\frac{2\pi\I\, \ell_h}{N_h}\right)\, R_h\,\Phi_g(x,z)\;,
\end{eqnarray}
\end{subequations}
where $h \in C_g$ has to be taken from the centralizer of $g$. We choose standard embedding 
eq.~\eqref{eqn:StandardEmbedding}, where the gauge embedding $R_g$ is identical to the geometrical 
action, i.e.\ $R_g = \theta$ for $g=(\theta, \lambda) \in S$, and the flavon $\Phi_g(x,z)$ is a 
triplet of $\SU{3}_\mathrm{fl}$. Note that the additional phases in eq.~\eqref{eqn:Delta54BC} (which 
dependent on $k$ and $\ell_h$, respectively) can originate from additional $\U{1}$ charges or from 
higher-dimensional Lorentz symmetry. These boundary conditions~\eqref{eqn:Delta54BC} result in the 
following conditions on the VEV of the zero mode $\Phi_g(x)$,
\begin{subequations}
\begin{eqnarray}
\braket{\Phi_g} & = & \exp\left(\frac{2\pi\I\, k}{N_g}\right)\, R_g\,\braket{\Phi_g}\;,\label{eqn:Delta54VEV1} \\
\braket{\Phi_g} & = & \exp\left(\frac{2\pi\I\, \ell_h}{N_h}\right)\, R_h\,\braket{\Phi_g}\;.\label{eqn:Delta54VEV2}
\end{eqnarray}
\end{subequations}

For each sector $g$ from table~\ref{tab:Delta54Sectors} we find some $k$ and $\ell_h$ such that the 
VEV $\braket{\Phi_g}$ is non-trivial. For example, consider the sector $g=(\vartheta,0)$ with 
$N_g=2$. Then, eq.~\eqref{eqn:Delta54VEV1} has two non-trivial solutions
\begin{subequations}
\begin{eqnarray}
\braket{\Phi_g} & = & v\,\left(\begin{array}{c}0\\ 1\\ -1\end{array}\right) \qquad\mathrm{for}\quad k=0\;,\\
\braket{\Phi_g} & = & \left(\begin{array}{c}v\\ w\\ w\end{array}\right) \;\;\quad\qquad\mathrm{for}\quad k=1\;,
\end{eqnarray}
\end{subequations}
for $v, w \in\mathbbm{C}$. Next, we have to ensure that these VEV alignments are invariant under 
transformations $h \in C_g$ from the centralizer. In this case, the centralizer $C_g$ is generated 
by $h=(\rho\vartheta\rho,0)$ which is of order $N_h = 6$. One can verify that 
eq.~\eqref{eqn:Delta54VEV2} has the same non-trivial solutions as before provided that $\ell_h$ 
takes some special values, i.e.\
\begin{subequations}
\begin{eqnarray}
\braket{\Phi_g} & = & v\,\left(\begin{array}{c}0\\ 1\\ -1\end{array}\right) \qquad\mathrm{for}\quad k=0 \quad\mathrm{and}\quad \ell_h=2\;,\\
\braket{\Phi_g} & = & \left(\begin{array}{c}v\\ w\\ w\end{array}\right) \;\;\quad\qquad\mathrm{for}\quad k=1 \quad\mathrm{and}\quad \ell_h=5\;.
\end{eqnarray}
\end{subequations}
We repeat this analysis for the other sectors of the $\Delta(54)$ orbifold, listed in 
table~\ref{tab:Delta54Sectors}, and find the following invariant VEV directions. The 
$\vartheta$-sector allows for two different boundary conditions that yield two flavon VEV alignments
\begin{subequations}
\begin{eqnarray}
\vartheta\braket{\phi}~=~\begin{pmatrix}-1&0&0\\0&0&-1\\0&-1&0\end{pmatrix}\braket{\phi} = \pm\braket{\phi}         &\to_{+}  & \braket{\phi}\propto\left(\begin{array}{c}0\\1\\-1\end{array}\right)\;,\\
                                                                                                                    &\to_{-}  & \braket{\phi}\propto\left(\begin{array}{c}a\\b\\ b\end{array}\right)\;.
\end{eqnarray}
\end{subequations}
For the $\omega$-sector we obtain three different VEV alignments
\begin{subequations}
\begin{eqnarray}
\omega\braket{\phi}   ~=~\begin{pmatrix} 0&1&0\\0&0& 1\\1& 0&0\end{pmatrix}\braket{\phi} = \alpha^{2n}\braket{\phi} &\to_{n=0}& \braket{\phi}\propto\left(\begin{array}{c}1\\1\\ 1\end{array}\right)\;,\\
                                                                                                                    &\to_{n=1}& \braket{\phi}\propto\left(\begin{array}{c}1\\ \alpha^2\\ \alpha^4\end{array}\right)\;,\\
                                                                                                                    &\to_{n=2}& \braket{\phi}\propto\left(\begin{array}{c}1\\ \alpha^4\\ \alpha^2\end{array}\right)\;,
\end{eqnarray}
\end{subequations}
while the $\rho$-sector yields
\begin{subequations}
\begin{eqnarray}
\rho\braket{\phi}     ~=~\begin{pmatrix} 0&\alpha^2&0\\0&0&\alpha^4\\1&0&0\end{pmatrix}\braket{\phi}=\alpha^{2n}\braket{\phi} &\to_{n=0}& \braket{\phi}\propto\left(\begin{array}{c}1\\ \alpha^4\\1\end{array}\right)\;,\\
                                                                                                                           &\to_{n=1}& \braket{\phi}\propto\left(\begin{array}{c}1\\ 1\\ \alpha^4\end{array}\right)\;,\\
                                                                                                                           &\to_{n=2}& \braket{\phi}\propto\left(\begin{array}{c}\alpha^4\\ 1\\ 1\end{array}\right)\;.
\end{eqnarray}
\end{subequations}
The matrices $\omega$ and $\rho$ are very similar only involving different phases, and we can only 
obtain one different matrix built from them
\begin{subequations}
\begin{eqnarray}
\omega^2\rho\braket{\phi}~=~\begin{pmatrix} 1&0&0\\0&\alpha^2&0\\0&0&\alpha^4\end{pmatrix}\braket{\phi}=\alpha^{2n}\braket{\phi} &\to_{n=0}& \braket{\phi}\propto\left(\begin{array}{c}1\\0\\0\end{array}\right)\;,\\
&\to_{n=1}& \braket{\phi}\propto\left(\begin{array}{c}0\\0\\1\end{array}\right)\,,\\
&\to_{n=2}& \braket{\phi}\propto\left(\begin{array}{c}0\\1\\0\end{array}\right)\;.
\end{eqnarray}
\end{subequations}
Multiplying $\omega$ to $\rho$ or $\vartheta$ would only cyclicly rotate the entries of the VEVS. 
The only other possibility would be to study the matrix
\begin{subequations}
\begin{eqnarray}
\omega^2\rho\vartheta\braket{\phi}~=~\begin{pmatrix} -1&0&0\\0&0&-\alpha^2\\0&\alpha^2&0\end{pmatrix}\braket{\phi}=\pm\braket{\phi} &\to_{+}& \braket{\phi}\propto\left(\begin{array}{c}0\\1\\-\alpha^4\end{array}\right)\;,\\
&\to_{-}& \braket{\phi}\propto\left(\begin{array}{c}a\\b\\ b\alpha^4\end{array}\right)\;.
\end{eqnarray}
\end{subequations}

We conclude that we can completely fix three flavons to have the TBM alignment choosing them to be eigenvectors of
\begin{equation}
\braket{\phi_1}=\omega\braket{\phi_1},\ \ \braket{\phi_2}=\vartheta\braket{\phi_2},\ \ \braket{\phi_3}=\alpha^2 \ \omega^2\rho\braket{\phi_3},
\end{equation}
while adding other matrices can introduce powers of $\alpha^2$ in any entry while keeping the same alignment.

\subsection{GUT breaking}
\label{sec:Delta54GUTBreaking}
As we stated before, in principle we could choose the gauge embedding $R_{e_i}$ of the 
translations to break the GUT, for example, to break $\SU{5}$. However, in this $\Delta(54)$ 
orbifold with standard embedding the consistency conditions force them to be unity, 
see~\eqref{eq:Delta54eiTrivial}. The simplest choice to avoid this situation in this orbifold would 
be to enlarge the $\mathbb{Z}_2$ generator
\begin{equation}
R_\vartheta ~=~ \begin{pmatrix} -1&0&0\\0&0&-1\\0&-1&0\end{pmatrix}\otimes P_{\SU{5}}\;,
\end{equation}
which is consistent with all conditions and breaks $\SU{5}$. Our model outlined in 
section~\ref{model} is based on this GUT breaking.

\subsection{SM fermion localization}

Quarks and leptons that transform as triplets under the $\SU{3}_\mathrm{fl}$ flavour symmetry 
should not feel any flavour breaking boundary conditions. Otherwise, some of them would be 
projected out by the orbifold and, hence, we would lose the predictivity from the flavon VEV 
alignments.

We can see from table~\ref{tab:Delta54Sectors} that this specific $\Delta(54)$ orbifold has 
specific locations with zero-dimensional fixed points. Any field localized at such a point in extra 
dimensions is already a 4d field and, hence, is not be subject to any boundary condition. 
Consequently, we localize our $\SU{3}_\mathrm{fl}$ lepton triplet at such a fixed point.

\section[SU(5) x SU(3)fl model in R4 x T6/Delta(54)]{\boldmath$\SU{5}\times \SU{3}_\mathrm{fl}$ model in $\mathbb{R}^4\times\mathbb{T}^6/\Delta(54)$\unboldmath}
\label{model}

We start with $\mathcal{N}=1$ SUSY in 10 dimensions and with a 
$\SU{5}_\mathrm{GUT}\times\SU{3}_\mathrm{fl}$ gauge symmetry. In addition, we impose a $\U{1}$ 
shaping symmetry that allows the required Yukawa sector. Then, we define the $\Delta(54)$ orbifold 
boundary conditions as
\begin{equation}\begin{split}
R_\vartheta ~=~ \begin{pmatrix} -1&0&0\\0&0&-1\\0&-1&0\end{pmatrix}\otimes P_{\SU{5}}, &\qquad R_\omega ~=~ \begin{pmatrix} 0&1&0\\0&0&1\\1&0&0\end{pmatrix}\otimes \Id_{5\times 5}, \\   
R_\rho ~=~ \begin{pmatrix} 0&\alpha^2&0\\0&0&\alpha^4\\1&0&0\end{pmatrix}\otimes\Id_{5\times 5}\;,&\qquad R_{e_i}=\Id_{3\times 3}\otimes\Id_{5\times 5},
\end{split}\end{equation}
where $\alpha = \exp(\frac{2\pi\I}{6})$ and $P_{\SU{5}}=\mathrm{diag}(1,1,-1,-1,-1)$. Since the embedding 
$R$ acts as standard embedding on $\SU{3}_\mathrm{fl}$, one can check easily that these matrices 
fulfil all the necessary conditions of the $\mathbb{T}^6/\Delta(54)$ orbifold.

As also discussed in section~\ref{sec:Delta54GUTBreaking} these boundary conditions break 
$\SU{5}_\mathrm{GUT}\times \SU{3}_\mathrm{fl}\to \SU{3}_\mathrm{C}\times\SU{2}_\mathrm{L}\times\U{1}_Y$ 
with only the MSSM superfields and some pure flavons left after compactification.

The list of chiral superfield is given in table~\ref{tab:fieldssu5}. There, the $\pm$ superscript 
indicates that there are two copies of each ten-plet $T_i$, $i=1,2,3$, i.e.\ one for each parity 
under the boundary condition $P_{\SU{5}}$. 

\begin{table}
\centering
\footnotesize
\captionsetup{width=0.9\textwidth}
\begin{tabular}{| c | c@{\hskip 5pt}c |  c |c| c| c|}
\hline
\multirow{2}{*}{\rule{0pt}{4ex}Field}	& \multicolumn{3}{c |}{Representation} & \multirow{2}{*}{\rule{0pt}{4ex}Localization}&\multirow{2}{*}{\rule{0pt}{4ex}extra dim.}	&\multirow{2}{*}{\rule{0pt}{4ex}Zero mode}	 \\
\cline{2-4}
\rule{0pt}{3ex}		& $\SU{3}_\mathrm{fl}$ & $\SU{5}_\mathrm{GUT}$ &  $\U{1}$ & & &\\ [0.75ex]
\hline \hline
\rule{0pt}{3ex}%
$F$ 			& 3& $\bar{5}$ &11 & $ \big[ \vartheta \rho \big]$& 0 & $d^c,\ \ell$\\
$T_1^+$ 		& 1& $10$      & 6 & $ \big[ \vartheta  \big]$& 2 & $u_1^c,\ e_1^c$\\
$T_1^-$ 		& 1& $10$      & 6 & $-\big[ \vartheta  \big]$& 2 & $q_1$\\
$T_2^+$ 		& 1& $10$      & 4 & $ \big[ \vartheta  \big]$& 2 & $u_2^c,\ e_2^c$\\
$T_2^-$ 		& 1& $10$      & 4 & $-\big[ \vartheta  \big]$& 2 & $q_2$\\
$T_3^+$ 		& 1& $10$      & 2 & $ \big[ \vartheta  \big]$ & 2& $u_3^c,\ e_3^c$\\
$T_3^-$ 		& 1& $10$      & 2 & $-\big[ \vartheta  \big]$& 2 & $q_3$\\
$N_a$ 			& 1& $1$       & 1 & $ \big[ \omega \big]$  & 2   & $n_a$\\
$N_s$ 			& 1& $1$       & 2 & $ \big[ \vartheta \big]$& 2  & $n_s$\\
$H_5$ 			& 1& $5$       &-4 & $ \big[ \Id \big]$      & 6  & $h_u$\\
$H_{\bar{5}}$ 		& 1& $\bar{5}$ &11 & $ \big[ \Id \big]$   & 6     & $h_d$\\
\hline
$\xi$ 			& 1& 1 & -2 & $\big[ \Id \big]$& 6 & $\xi^0$\\
$\xi'$ 			& 1& 1 & -2 & $\big[ \vartheta \rho \big]$& 0& $\xi'^0$\\
\hline
$\phi_s$ 		& $\bar{3}$& 1  & -7& $\big[ \omega \big]$& 2 &$\phi_s^0 \propto (1,1,1)^\mathrm{T}$ \\
$\phi_a$ 		& $\bar{3}$& 1  &  -8& $\big[ \vartheta \big]$& 2&$\phi_a^0\propto (0,1,-1)^\mathrm{T}$\\
$\phi_\tau$ 		& $\bar{3}$& 1 & -2 &$\alpha^2\big[ \omega^2 \rho \big]$& 2 &$\phi_\tau^0\propto (0,0,1)^\mathrm{T}$\\
$\phi_\mu$ 		& $\bar{3}$& 1 & -4 &$\alpha^4\big[ \omega^2 \rho \big]$ & 2&$\phi_\mu^0\propto (0,1,0)^\mathrm{T}$\\
$\phi_e$ 		& $\bar{3}$& 1 & -6 &$\big[ \omega^2 \rho \big]$& 2 &$\phi_e^0\propto (1,0,0)^\mathrm{T}$\\
\hline
\end{tabular}
\caption{Complete list of chiral superfields in the model.  The $\U{1}$ is a shaping symmetry.}
\label{tab:fieldssu5}
\end{table}

We assume a standard K\"ahler potential with canonical normalized fields (and without large 
corrections~\cite{Chen:2013aya}). Then, the Yukawa sector after compactification reads
\begin{equation}
\begin{split}
\mathcal{W}_Y&     =Y_{ij}^u\, h_u\, q_i\, u_j^c \\
             &\quad+\frac{y_{33}^+}{\Lambda}                                                  h_{d} \left(\ell\cdot\phi_\tau^0\right) e_3^c 
                   +\frac{y_{22}^+}{\Lambda}                                                  h_{d} \left(\ell\cdot\phi_\mu^0\right)  e_2^c
                   +\frac{y_{11}^+}{\Lambda}                                                  h_{d} \left(\ell\cdot\phi_e^0\right)    e_1^c\\ 
             &\quad+\frac{y_{23}^+ \tilde\xi}{\Lambda}                                        h_{d} \left(\ell\cdot\phi_\tau^0\right) e_2^c 
                   +\frac{y_{13}^+ \tilde\xi^2 + y_{13}'^+ \tilde\xi'^2}{\Lambda}             h_{d} \left(\ell\cdot\phi_\tau^0\right) e_1^c 
                   +\frac{y_{12}^+ \tilde\xi}{\Lambda}                                        h_{d} \left(\ell\cdot\phi_\mu^0\right)  e_1^c\\
             &\quad+\frac{y_{33}^-}{\Lambda}                                                  h_{d} \left(d^c\cdot\phi_\tau^0\right) q_3 
                   +\frac{y_{22}^-}{\Lambda}                                                  h_{d} \left(d^c\cdot\phi_\mu^0\right)  q_2
                   +\frac{y_{11}^-}{\Lambda}                                                  h_{d} \left(d^c\cdot\phi_e^0\right)    q_1\\ 
             &\quad+\frac{y_{23}^- \tilde\xi}{\Lambda}                                        h_{d} \left(d^c\cdot\phi_\tau^0\right) q_2 
                   +\frac{y_{13}^- \tilde\xi^2 + y_{13}'^- \tilde\xi'^2}{\Lambda}             h_{d} \left(d^c\cdot\phi_\tau^0\right) q_1 
                   +\frac{y_{12}^- \tilde\xi}{\Lambda}                                        h_{d} \left(d^c\cdot\phi_\mu^0\right)  q_1\\
             &\quad+y^N_a \xi                                                                   n_a n_a
                   +\frac{y^N_s \xi^2 +y'^N_s \xi'^2}{\Lambda}                                  n_s n_s 
                   +\frac{y_a^\nu}{\Lambda}                                                     h_u \left(\ell\cdot\phi_a^0\right)    n_a
                   +\frac{y_s^\nu \tilde\xi}{\Lambda}                                           h_u \left(\ell\cdot\phi_s^0\right)    n_s\\
             &\quad+\frac{y_e^\nu \tilde\xi'}{\Lambda}                                          h_u \left(\ell\cdot\phi_e^0\right)    n_a 
                   +\frac{y_\mu^\nu \tilde\xi \tilde\xi'}{\Lambda}                              h_u \left(\ell\cdot\phi_\mu^0\right)  n_a 
                   +\frac{y_\tau^\nu \tilde\xi'\tilde\xi^2 + y_\tau'^\nu \tilde\xi'^3}{\Lambda} h_u \left(\ell\cdot\phi_\tau^0\right) n_a\;,
\label{eq:yuk}
\end{split}
\end{equation}
where
\begin{equation}
Y^u = \left(\begin{array}{ccc}y_{11}^u\,\tilde\xi^4 & y_{12}^u\,\tilde\xi^3 & y_{13}^u\,\tilde\xi^2 \\y_{21}^u\,\tilde\xi^3 & y_{22}^u\,\tilde\xi^2 & y_{23}^u\,\tilde\xi \\y_{31}^u\,\tilde\xi^2 & y_{32}^u\,\tilde\xi & y_{33}^u \end{array}\right)
    + \tilde\xi'^2 \left(\begin{array}{ccc}y_{11}'^u\,\tilde\xi^2 & y_{12}'^u\,\tilde\xi & y_{13}'^u \\y_{21}'^u\,\tilde\xi & y_{22}'^u & 0 \\y_{31}'^u & 0 & 0 \end{array}\right)
    + \tilde\xi'^4 \left(\begin{array}{ccc}y_{11}''^u & 0 & 0 \\0 & 0 & 0 \\0 & 0 & 0 \end{array}\right)\;.
\end{equation}
We have defined
\begin{equation}
\tilde{\xi} ~=~ \braket{\xi}/\Lambda \quad\mathrm{and}\quad \tilde{v}_i ~=~ \braket{v_i}/\Lambda\;.
\label{scaled}
\end{equation}
The $\U{1}$ shaping symmetry allows only these terms and there are no higher order contributions.

Note that the terms coming from $H_5 T_i^- T_j^+$ and $H_{\bar{5}}F\phi_k T_i^\pm \xi^\ell$ satisfy the basic 
string selection rule for allowed interactions: the point group selection rule demands that the 
point group elements of the respective constructing elements multiply to the identity element. For 
example, $H_{\bar{5}}$ and $\xi$ originate from the bulk with constructing element $\big[ \Id \big]$, $F$ 
is localized in the $\big[ \vartheta \rho \big]$ sector, the fields $\phi_k$ for $k=\tau, \mu, e$ 
live in the $\big[ \omega^2 \rho \big]$ sector and, finally, the fields $T_i^\pm$ for $i=1,2,3$ 
are localized in the $\big[ \vartheta  \big]$ sector. Then,
\begin{equation}
\big[ \Id \big]\; \big[ \vartheta \rho \big]\; \big[ \omega^2 \rho \big]\; \big[ \vartheta  \big]\; \big[ \Id \big]^\ell ~\supset~ \Id\;  \vartheta\omega\rho^2\; \rho\omega^2 \; \vartheta \; \Id^\ell ~=~ \Id\;,
\end{equation}
using the conjugacy classes of $\Delta(54)$ given in eq.~\eqref{eqn:Delta54ConjugacyClasses}.

We assume all dimensionless couplings to be $\mathcal{O}(1)$ complex numbers, so that all the hierarchies are due to the flavon VEVs
\begin{equation}
1000\ \tilde v_s\sim 1000\ \tilde v_a\sim 100\ \tilde v_e\sim 10\ \tilde v_\mu\sim 10\ \tilde\xi'\sim\tilde v_\tau\sim\tilde{\xi}\sim 0.1,
\label{eqn:vevhierarchies}
\end{equation}
which is an assumption.

With these assumptions we may approximate $\tilde\xi+\tilde\xi'\approx \tilde\xi$.
We write the up quark mass matrix, coming from the first line of eq. \ref{eq:yuk}  \footnote{All the mass matrices are given in the LR convention.}
\begin{equation}
M^u=v_u\left(\begin{array}{ccc}
y_{11}^u\tilde{\xi}^4 & y_{12}^u\tilde{\xi}^3 & y_{13}^u\tilde{\xi}^2\\
y_{21}^u\tilde{\xi}^3 & y_{22}^u\tilde{\xi}^2 & y_{23}^u\tilde{\xi}\\
y_{31}^u\tilde{\xi}^2 & y_{32}^u\tilde{\xi} & y_{33}^u
\end{array}\right).
\label{eq:up_mass}
\end{equation}

The next lines of eq. \ref{eq:yuk} give masses to down quarks and charged leptons. The down quark matrix is
\begin{equation}
M^d=v_d\left(\begin{array}{ccc}
y_{11}^-\tilde{v}_e & y_{12}^-\tilde{v}_\mu\tilde{\xi} & y_{13}^-\tilde{v}_\tau\tilde{\xi}^2\\
0 & y_{22}^-\tilde{v}_\mu & y_{23}^-\tilde{v}_\tau\tilde{\xi}\\
0 & 0 & y_{33}^-\tilde{v}_\tau
\end{array}\right),
\label{eq:down_mass}
\end{equation}
while the charged lepton mass matrix is
\begin{equation}
(M^e)^*=v_d\left(\begin{array}{ccc}
y_{11}^+\tilde{v}_e & 0&0 \\
y_{12}^+\tilde{v}_\mu\tilde{\xi}  & y_{22}^+\tilde{v}_\mu & 0\\
y_{13}^+\tilde{v}_\tau\tilde{\xi}^2 & y_{23}^+\tilde{v}_\tau\tilde{\xi} & y_{33}^+\tilde{v}_\tau
\end{array}\right).
\label{eq:charged_mass}
\end{equation}
Since $e^c$ comes from $T^+$ and $q$ comes from $T^-$ the Yukawa terms have different and independent couplings $y^\pm_{ij}$ for each one. This way the charged lepton mass matrix is completely independent of the down quark mass matrix.

The final two lines in eq. \ref{eq:yuk} give the Dirac neutrino mass matrix and the right handed neutrino Majorana mass matrix
\begin{equation}
M^\nu_D=v_u\left(\begin{array}{cc} {y}^\nu_e\tilde{v}_e\tilde{\xi}' & y_s^\nu \tilde{v}_s\tilde{\xi}\\
y_a^\nu \tilde{v}_a+ {y}^\nu_\mu \tilde{v}_\mu\tilde{\xi}\tilde\xi'& y_s^\nu \tilde{v}_s\tilde{\xi}\\
-y_a^\nu \tilde{v}_a+ \tilde{v}_\tau({y}^\nu_\tau\tilde{\xi}^2\tilde\xi'+{y}'^\nu_\tau\tilde\xi\tilde\xi'^2)& y_s^\nu \tilde{v}_s
\tilde{\xi} \end{array}\right), \ \ \ 
M^N =\left(\begin{array}{cc}
y^N_a\tilde{\xi}&0 \\ 0 &y^N_s\tilde{\xi}^2\\
\end{array}\right)\braket{\xi}.
\end{equation}

From the assumed VEV hierarchies from eq. \ref{eqn:vevhierarchies}, the $\tilde\xi$ terms in the first column are expected to be one order of magnitude smaller than the ones without $\tilde{\xi}$, so we may safely ignore them, leading to 
\begin{equation}
M^\nu_D\simeq v_u\left(\begin{array}{cc} {y'}^\nu_e\tilde{v}_e\tilde{\xi}' & y_s^\nu \tilde{v}_s\tilde{\xi}\\
y_a^\nu \tilde{v}_a& y_s^\nu \tilde{v}_s\tilde{\xi}\\
-y_a^\nu \tilde{v}_a& y_s^\nu \tilde{v}_s\tilde{\xi}
\end{array}\right)
\sim \left(\begin{array}{cc} \epsilon & b \\
a & b \\
- a & b 
\end{array}\right)\;.
\end{equation}
In the limit that the small entry denoted by $\epsilon$ is ignored, the Dirac mass matrix is of the 
CSD form and leads to tribimaximal neutrino mixing~\cite{King:2005bj}. The presence of $\epsilon$ 
has the effect of switching on the reactor angle $\theta_{13}$, without modifying very much the 
solar and atmospheric angles from their tribimaximal values~\cite{Harrison:2002er}. This corresponds to so called 
tribimaximal-reactor lepton mixing~\cite{King:2009qt}.

The RHN are very heavy so that the left handed neutrinos become very light after the Seesaw mechanism has been implemented,
\begin{equation}\begin{split}
M^\nu &=M^\nu_D(M^N)^{-1}(M^\nu)^\mathrm{T},\\
M^\nu \frac{\braket{\xi}}{v_u^2}&\simeq\frac{(y^\nu_a)^2\tilde{v}_a^2}{y^N_a}\left(\begin{array}{ccc} (y_{12}^\nu\tilde{\xi}')^2 &y_{12}^\nu\tilde{\xi}' &-y_{12}^\nu\tilde{\xi}'  \\ y_{12}^\nu\tilde{\xi}'  & 1 & -1 \\- y_{12}^\nu\tilde{\xi}'  & -1 &1\end{array}\right)+\tilde{\xi}\frac{(y^\nu_s)^2\tilde{v}_s^2}{y^N_s}\left(\begin{array}{ccc} 1 &1&1 \\ 1 & 1 & 1 \\ 1 & 1 &1\end{array}\right),
\end{split}
\label{eq:nu_mass}
\end{equation}
where $y_{12}^\nu={y}_e^\nu \tilde{v}_e/y_a^\nu \tilde{v}_a$.

Looking at all the mass matrices, as noted above, we may see that the VEV $\tilde{\xi}'$ only 
appears in the neutrino mass matrix. Knowing that the charged lepton correction to the PMNS are 
negligible, if we sent this $\tilde{\xi}'\to 0$, we would have the tribimaximal (TBM) setup for the 
PMNS. Therefore the sole role of the $\tilde{\xi}'$ is to deviate from the 
TBM~\cite{King:2009qt,King:2012vj,King:2011ab}.

\subsection{Numerical fit in the neutrino sector}
In this section we perform a fit to the PMNS observables and the neutrino masses assuming a diagonal charged-lepton mass matrix (the off-diagonal elements in eq.~\ref{eq:charged_mass} are negligible due to the appearance of $\tilde{\xi}$). The complex parameters in the up-type and down-type quark mass matrices in eqs.~\ref{eq:up_mass} and~\ref{eq:down_mass} have enough freedom to fit all the quark masses and the observed CKM mixing angles. Therefore, in this section we focus only on the neutrino sector. 

 The effective neutrino mass matrix in eq.~\ref{eq:nu_mass} can be rewritten in terms of input parameters as 
\begin{equation}
M^\nu \simeq \mu_a \left(\begin{array}{ccc} (\tilde{\xi}_{12} e^{i\eta_{\tilde{\xi}}})^2 &\tilde{\xi}_{12} e^{i\eta_{\tilde{\xi}}} &-\tilde{\xi}_{12} e^{i\eta_{\tilde{\xi}}}  \\ \tilde{\xi}_{12} e^{i\eta_{\tilde{\xi}}}  & 1 & -1 \\- \tilde{\xi}_{12} e^{i\eta_{\tilde{\xi}}}  & -1 &1\end{array}\right)+ \mu_s e^{i \eta}\left(\begin{array}{ccc} 1 &1&1 \\ 1 & 1 & 1 \\ 1 & 1 &1\end{array}\right),
\label{eq:nu_mass_fit}
\end{equation}
where we have defined the input parameters
\begin{equation}\begin{split} \mu_a&=\left|\frac{v_u^2}{\braket{\xi}} \frac{(y^\nu_a)^2\tilde{v}_a^2}{y^N_a} ,\right| \quad  \tilde{\xi}_{12}=|y^\nu_{12}\tilde{\xi}'|, \quad  \eta_{\tilde{\xi}}=\arg{(y^\nu_{12}\tilde{\xi})}, \\ \mu_s&=\left|\frac{v_u^2}{\braket{\xi}}\tilde{\xi}\frac{(y^\nu_s)^2\tilde{v}_s^2}{y^N_s}\right| \quad \mathrm{and} \quad   \eta=\arg\left(\frac{(y^\nu_s)^2\tilde{v}_s^2\tilde\xi}{y^N_s} \frac{y^N_a}{(y^\nu_a)^2\tilde{v}_a^2}\right)
\end{split}
\label{eq:input_parameters}
\end{equation}
We implement a numerical fit using a $\chi^2$ test function
\begin{equation}
\chi^2=\sum_n \left(\frac{P_n(x)-P_n^\mathrm{obs}}{\sigma_n}\right)^2,
\end{equation}	
where we sum over the 6 observables given by $P_n^\mathrm{obs}=\{ \theta^\ell_{12},\theta^\ell_{13},\theta^\ell_{23}, \delta^l,  \Delta m_{21}^2, \Delta m_{31}^2  \}$ with statistical errors $\sigma_n$. The predictions of the model for these observables are given by $P_n(x)$, where $x=\{\mu_a,\tilde{\xi}_{12},\eta_{\tilde{\xi}},\mu_s, \eta\}$ refers to the different input parameters. We are doing the numerical fit in terms of the effective neutrino mass matrix in eq.~\eqref{eq:nu_mass_fit} and we ignore any renormalisation group running corrections.

We use the recent global fit values of neutrino data from NuFit4.1~\cite{Esteban:2018azc}. Most of the observables follow an almost Gaussian distribution and we take a conservative approach using the smaller of the given uncertainties in our computations except for $\theta^l_{23}$ and $\delta^l$. The best fit from NuFit4.1 is for normal mass ordering with inverted ordering being disfavoured with a $\Delta\chi^2=6.2 (10.4) $ without (with) the Super-Kamiokande atmospheric neutrino data analysis.

The model predictions are shown in table~\ref{tab:numleptons}. The neutrino mass matrix in eq.\ref{eq:nu_mass_fit} predicts near maximal atmospheric mixing angle $\theta_{23}^l=45.66^\circ$ and solar mixing angle $\theta_{12}=35^\circ$ as expected since we only have a small correction from tribimaximal mixing which allows a correct non-zero reactor angle $\theta_{13}^l=8.615^\circ$. The  CP violation prediction is $\delta^l\approx 225^\circ$. All the model predictions for the PMNS observables and the neutrino mass-squared differences are within the $3\sigma$ region from the latest neutrino oscillation data and reproduce a $\chi^2=4.8$ value. Furthermore, since we only have 2RH neutrinos, $m_1=0$ and there is only one physical Majorana phase $\alpha_{23}$~\cite{Tanabashi:2018oca}. The bound on the effective Majorana mass $m_{\beta\beta}$ is taken from~\cite{Agostini:2018tnm} while the prediction is also given in table~\ref{tab:numleptons}.
\begin{table}[!ht]
	\centering
	\footnotesize
	\renewcommand{\arraystretch}{1.1}
	\begin{tabular}{ l cc c c }
		\hline
		\multirow{2}{*}{Observable} & \multicolumn{2}{c}{Data} && \multicolumn{1}{c}{Model} \\
		\cline{2-5} 
		& Central value & $3\sigma$ range && Best fit \\
		\hline
		$\theta_{12}^l$ $/^\circ$ &33.82 & 31.61 $\to$ 36.27 && 35.00  \\	
		$\theta_{13}^l$ $/^\circ$ &8.610 & 8.220 $\to$ 8.990 && 8.615  \\
		$\theta_{23}^l$ $/^\circ$ &48.30& 40.80 $\to$ 51.30 &&45.66 \\	
		$\delta^l$ $/^\circ$ & 222.0 & 141.0 $\to$ 370.0 && 225.3 \\
		$\Delta m_{21} ^2$  $/ 10^{-5}\mathrm{eV}^2$ & 7.390& 6.790 $\to$ 8.010 &&7.393  \\	
		$\Delta m_{31}^2$  $/ 10^{-3}\mathrm{eV}^2$ & 2.523 & 2.432 $\to$ 2.618 && 2.525 \\
		$m_1$ /meV & & &&  0 \\ 
		$m_2$ /meV & & &&  8.599 \\ 
		$m_3$ /meV & & &&  50.25\\
		$\sum m_i$ /meV & \multicolumn{2}{c}{$\lesssim$ 230 } &&  58.85\\
		$ \alpha_{23} $ $/^\circ$ & & && 221.3 \\
		$ m_{\beta\beta}$ /meV &  \multicolumn{2}{c}{$\lesssim$ 60-200}   && 2.754\\
		\hline
	\end{tabular}
	\caption{Latest values of PMNS observables and neutrino masses given by NuFit4.1~\cite{Esteban:2018azc} together with the model predictions with $\chi^2\approx 4.8$. The neutrino masses $m_i$ as well as the Majorana phase $\alpha_{23}$ are pure predictions of our model. The bound on $ \sum m_i $ is taken from~\cite{Ade:2015xua}.
		The bound on $ m_{\beta\beta}$  is taken from~\cite{Agostini:2018tnm}. 
		There is only one physical Majorana phase since $m_1=0$.}
	\label{tab:numleptons}
\end{table}

Table~\ref{tab:parameters} shows the input parameter values. There are 3 real parameters $\{\mu_a, \tilde{\xi}_{12}, \mu_s\}$ plus two additional phases  $\{\eta_{\tilde{\xi'}}, \eta \}$, a total of 4 input parameters to fit 6 data points.
Naively, we can measure the goodness of the fit computing the reduced $\chi^2$, i.e. the $\chi^2$ per degree of freedom $\chi^2_\nu=\chi^2/\nu$. The number of degrees of freedom is given by $\nu=n-n_i$, where $n=6$ is the number of measured observables, while $n_i=4$ is the number of input parameters. A good fit is expected to have $\chi^2_\nu\sim1$. We have 2 degrees of freedom and the best fit has a reduced $\chi^2_\nu \simeq 2.4$. We view this as a good fit and it also remarks the predictivity of the model, not only fitting to all available quark and lepton data but also fixing the neutrino masses and Majorana phases. 

\begin{table}[ht]
	\centering
	\footnotesize
	\renewcommand{\arraystretch}{1.1}
	\begin{tabular}[t]{lr}
		\hline
		Parameter & Value \\ 
		\hline
		$\mu_a / 10^{-2}\mathrm{meV} $ & $2.454$ \\
		$\tilde{\xi}_{12} $ & $0.221$ \\
		$\eta_{\tilde{\xi'}}$ & $-2.392$ \\
		$\mu_s / 10^{-3}\mathrm{meV} $ &2.887 \\
		$\eta $ & $-2.422$ \\
		\hline
	\end{tabular}
	\caption{Input parameter values reproducing the best fit point with $\chi^2 \sim 4.8$.} 
	\label{tab:parameters}
\end{table}

Using the definition of the input parameters in eq.~\ref{eq:input_parameters} and their values in table~\ref{tab:parameters} for the best fit point, we can give a naive estimation of the value of $\tilde{\xi}$. If we assume the dimensionless parameters to be $\mathcal{O}(1)$ and the VEVs $\tilde{v}_a \sim \tilde{v}_s$, then we find $\mu_s/\mu_a \approx \tilde{\xi} \approx 0.1
$, which justifies the assumption of an approximate diagonal charged-lepton matrix in eq.~\ref{eq:charged_mass} and the values of the VEVs in eq.~\ref{eqn:vevhierarchies}. Also note that $|\tilde{\xi}_{12}|\approx 0.22$ which is exactly the value of Cabbibo angle~\cite{King:2009qt,King:2012vj,King:2011ab}.

\section{Conclusions}
\label{sec:conclusions}

The flavour puzzle, in particular the large mixing observed in the lepton sector, provides a strong motivation for going beyond the Standard Model. The literature is replete with flavour models involving some family symmetry spontaneously broken by flavon fields with certain vacuum alignments motivated by phenomenological considerations, but highly non-trivial to achieve without resort to extra symmetries and driving fields. In order to overcome this obstacle, one promising approach is to attempt to formulate such theories in extra dimensions, where the desired vacuum alignments may emerge from orbifold boundary conditions.

We have systematically developed the formalism necessary for ensuring that boundary conditions of 
flavon fields in extra dimensions are consistent with heterotic string theory. Having developed a 
set of consistency conditions on the boundary conditions, we have then explored a series of 
examples of orbifolds in various dimensions to see which ones can satisfy them. In addition we have 
imposed the further phenomenological requirement of having non-trivial flavon vacuum alignments. 
We have also demanded that quarks and leptons be located appropriately in extra dimensions so that 
their massless modes may include complete multiplets under the gauged flavour symmetry. 

It turns out that it is highly non-trivial to satisfy all of these conditions (theoretical and 
phenomenological) together. For instance, the simple $\mathbb{T}^2/\mathbb{Z}_2$ orbifold, while 
allowing SM fermion matter localisation on fixed points, does not permit non-trivial flavon vacuum 
alignments, consistently with the formal requirements of the boundary conditions. This motivates us 
to go to 10d models. However, the simple orbifold $\mathbb{T}^6/(\Z{2}\times\Z{2})$ fares no better 
than the previous case, since it too can only provide one non-trivial alignment. We find that the 
boundary conditions must exhibit some non-Abelian structure so that we can have non-trivial VEV 
alignments. 

Following the above logic, we were led to consider 10d non-Abelian orbifolds, where the torus is 
modded out by a non-Abelian group. We have studied the orbifold $\mathbb{T}^6/S_4$ which can 
fix flavons into the highly predictive CSDn structure. This orbifold however, does not have 4d 
branes where the SM matter could be localized. This motivated us to consider the orbifold 
$\mathbb{T}^6/\Delta(54)$, as an example where we can locate the SM fermions on fixed points in 
extra dimensions. Since the $\Delta(54)$ orbifold is not so well studied in the literature, we have 
developed this case in some detail, and eventually shown that we can choose the extra dimensions in 
such a way that we can build a realistic model. 

The minimal successful flavour theory seems to be a 10d theory with a $\SU{3}_\mathrm{fl}$ gauged 
flavour symmetry, where the six extra dimensions are compactified on a $\mathbb{T}^6/\Delta(54)$ 
orbifold. The $\SU{3}_\mathrm{fl}$ flavour symmetry is broken by flavon VEVs which are completely 
aligned by the boundary conditions of the orbifold. The vacuum alignment of the flavons is of the 
tribimaximal form, but the theory can allow for some small corrections leading to 
tribimaximal-reactor lepton mixing, which we have shown to be consistent with current neutrino data.
We have constructed a fully realistic $\SU{5}\times \SU{3}_\mathrm{fl}$ grand unified theory along 
these lines, which is complete, predictive and in principle consistent with heterotic string 
theory.

\section*{Acknowledgments}
SFK acknowledges the STFC 
Consolidated Grant ST/L000296/1 
and InvisiblesPlus RISE No.\ 690575. 
SFK and EP acknowledge the European Union's Horizon 2020 
Research and Innovation programme under Marie Sk\l {}odowska-Curie grant 
agreement Elusives ITN No.\ 674896. PV is supported by the Deutsche Forschungsgemeinschaft (SFB1258).

\appendix
\section{Definition of an orbifold by its space group}
\label{app:SpaceGroup}

In order to define a $D$-dimensional toroidal orbifold $\mathbbm{O}$ geometrically as a quotient 
space of $D$-dimensional space, i.e.\
\begin{equation}\label{eqn:geometricorbifold}
\mathbbm{O} = \mathbbm{R}^D/S\;,
\end{equation}
one has to specify a space group $S$ first. A general element $g$ of a space group $S$ consists of a 
rotation $\theta \in \SO{D}$ (also called twist) and a translation $\lambda \in \mathbbm{R}^D$, i.e.
\begin{equation}
g ~=~ (\theta, \lambda) ~\in~ S\;.
\end{equation}
By definition, $g$ acts on the internal coordinates $z \in \mathbbm{R}^D$ as
\begin{equation}
z ~\stackrel{g}{\longmapsto}~ g\; z ~=~ (\theta, \lambda)\; z ~=~ \theta\,z + \lambda\;.
\end{equation}
Consequently, two space group elements $g_1 = (\theta_1, \lambda_1) \in S$ and 
$g_2 = (\theta_2, \lambda_2) \in S$ multiply as
\begin{equation}
g_1\, g_2 ~=~ (\theta_1, \lambda_1)\, (\theta_2, \lambda_2) ~=~ (\theta_1\, \theta_2, \theta_1\,\lambda_2 + \lambda_1)\;.
\end{equation}
Furthermore, the inverse element $g^{-1} \in S$ of $g = (\theta, \lambda) \in S$ is given by
\begin{equation}
g^{-1} ~=~ (\theta^{-1}, -\theta^{-1}\,\lambda)\;,
\end{equation}
and the neutral element is
\begin{equation}
g_\Id ~=~ (\Id, 0) ~\in~ S\;.
\end{equation}
Hence, one can see that $S$ is a discrete group, actually a discrete subgroup of the 
extra-dimensional Euclidean group.

Practically, one defines a space group by a (finite) list of generators, which are pure 
translations and rotations. In this work, we focus on the case of up to three rotational 
generators\footnote{Ignoring the possibility of roto-translations, i.e.\ rotations that are combined 
with fractional translations, for example $(\theta, \lambda)$ with $\lambda \not\in\Gamma$.}, i.e.
\begin{equation}
(\Id, e_i) \quad , \quad (\vartheta, 0) \quad , \quad (\omega, 0)\quad , \quad (\rho, 0)\;,
\end{equation}
for $i\in\{1,\ldots,D\}$. The vectors $e_i \in \mathbbm{R}^D$ span a $D$-dimensional lattice $\Gamma$ 
that specifies a $D$-dimensional torus $\mathbbm{T}^D = \mathbbm{R}^D/\Gamma$ and the rotations 
$\vartheta$, $\omega$ and $\rho$ have to be symmetries of the lattice $\Gamma$, i.e.
\begin{equation}
\vartheta\, \Gamma ~=~ \Gamma \;\;,\;\; \omega\, \Gamma ~=~ \Gamma \quad\mathrm{and}\quad \rho\, \Gamma ~=~ \Gamma\;.
\end{equation}
The rotations $\vartheta$, $\omega$ (and $\rho$) generate the so-called point group $P$, where we are 
dealing with the cases $P = \Z{2}\times\Z{2}$ in section~\ref{sec:Z2xZ2Orbifold}, $P = S_4$ in 
section~\ref{sec:S4Orbifold} and $P = \Delta(54)$ in section~\ref{sec:Delta54}. 

Having defined a space group, the orbifold $\mathbbm{O}$ given in eq.~\eqref{eqn:geometricorbifold} 
is defined by identifying those points $z_{(1)}$ and $z_{(2)}$ in $\mathbbm{R}^D$ that are mapped 
to each other by some element of the space group, i.e.
\begin{equation}\label{eqn:GeometricalOrbifold}
z_{(1)} ~\sim~ z_{(2)} \quad \Leftrightarrow~ \mathrm{there\ is }\ g \in S\ \mathrm{ such\ that }\ z_{(1)} = g\, z_{(2)}\;.
\end{equation}
This equivalence relation can be used to define a fundamental domain of the orbifold.

A given element $g \in S$ of the space group can have a set of fixed points $\mathrm{F}_g$, defined by
\begin{equation}
\mathrm{F}_g ~:=~ \{z \in \mathbbm{R}^D ~|~ g\,z ~=~ z\}\;.
\end{equation}
For a given space group element $g=(\theta,\lambda) \in S$ (with appropriate translation $\lambda$) 
the dimension of $\mathrm{F}_g$ depends on the eigenvalues of the rotation matrix $\theta \in P$: 
Each eigenvalue +1 corresponds to an invariant direction in $\mathrm{F}_g$. Our main concern is the 
case of supersymmetric orbifolds in $D=6$, where we find fixed point sets of dimensions six (i.e.\ 
the bulk $\mathbbm{O}$ for $g=\Id$), two (i.e.\ fixed tori) and zero (i.e.\ fixed points). By acting with 
$h \in S$ onto the fixed point equation $g\,z = z$ (i.e.\ $z \in \mathrm{F}_g$), one obtains
\begin{equation}
\left(h\,g\,h^{-1}\right)\,\left(h\,z\right) ~=~ \left(h\,z\right)\;.
\end{equation}
Hence, $h\,z \in \mathrm{F}_{h\,g\,h^{-1}}$. However, due to eq.~\eqref{eqn:GeometricalOrbifold} 
points $z$ and $h\,z$ are identified on the orbifold $\mathbbm{O}$. Thus, the 
corresponding fixed point sets are identified on the orbifold $\mathbbm{O}$ as well,
\begin{equation}
\mathrm{F}_{g} ~\sim~ \mathrm{F}_{h\,g\,h^{-1}}\;,
\end{equation}
for all $h \in S$. Consequently, the inequivalent fixed point sets correspond to the 
conjugacy classes $[g]$ of $S$, where
\begin{equation}\label{eq:ConjugacyClasses}
[g] ~=~ \{ h\,g\,h^{-1} ~|~ h ~\in~ S\}\;.
\end{equation}
If the point group $P$ is Abelian, each element $\tilde{g}$ of a conjugacy class $[g]$ has the 
same point group element $\theta \in P$, i.e.
\begin{equation}
g ~=~ (\theta, \lambda) \quad\Leftrightarrow\quad h\,g\,h^{-1} ~=~ (\theta, \lambda') \;\ \mathrm{for\ all}\ h ~\in~ S\ \mathrm{and\ some}\ \; \lambda' ~\in~ \Gamma\;.
\end{equation}

\subsection{Orbifold-invariant fields}
\label{app:OrbifoldInvariantFields}

In this appendix we complete the discussion from section~\ref{sec:OrbifoldBoundaryConditions} in 
the case $g\, h \neq h\, g$. In this case, we can choose the proportionality in 
eq.~\eqref{eqn:IdentificationOfTwistedStrings} to be trivial, i.e.
\begin{equation}\label{eqn:IdentificationOfTwistedStrings2}
\Phi_g(x,h^{-1}z) ~=~ \Phi_{h\,g\,h^{-1}}(x,z)\;,
\end{equation}
where a possible phase has been absorbed in a redefinition of $\Phi_{h\,g\,h^{-1}}$. 
Consequently, all fields $\Phi_{h\,g\,h^{-1}}$ from the same conjugacy class 
$[g] = \{ h\,g\,h^{-1} |\, h \in S\}$ are identified 
and eqs.~\eqref{eqn:GeneralOrbifoldOfField2} and~\eqref{eqn:IdentificationOfTwistedStrings2} yield
\begin{equation}
\Phi_g(x,z) ~\stackrel{h}{\longmapsto}~ R_h\, \Phi_g(x,h^{-1}z) ~=~ R_h\, \Phi_{h\,g\,h^{-1}}(x,z)\;.
\end{equation}
Then, we can construct an orbifold-invariant field, denoted by $\Phi_{[g]}(x,z)$, on the covering 
space $\mathbbm{R}^D$ of $\mathbbm{O}$. To do so, we have to build the following linear combination
\begin{equation}\label{eqn:OrbifoldInvariantField}
\Phi_{[g]}(x,z) ~:=~ \Phi_g(x,z) + \sum_{h \,\not\in\, C_g}\, R_h\, \Phi_{h\,g\,h^{-1}}(x,z)\;,
\end{equation}
ignoring the normalization of $\Phi_{[g]}(x,z)$. However, $\mathrm{F}_g$ and 
$\mathrm{F}_{h\,g\,h^{-1}}$ are identified on the orbifold $\mathbbm{O}$. Hence, if we restrict 
$z \in \mathbbm{O}$ (instead of $z \in \mathbbm{R}^D$) we can ignore the contributions 
$\Phi_{h\,g\,h^{-1}}(x,z)$ in eq.~\eqref{eqn:OrbifoldInvariantField} and use $\Phi_g(x,z)$ as a 
well-defined field on the orbifold $\mathbbm{O}$. In this case, 
transformations~\eqref{eqn:GeneralOrbifoldOfField2} with $h \not\in C_g$ are not considered as they 
would map a point $z$ from the fundamental domain of the orbifold to a point outside of the 
fundamental domain.

\section[Details on the S4 orbifold]{\boldmath Details on the $S_4$ orbifold\unboldmath}
\label{app:S4OrbifoldDetails}

Beside the untwisted sector with constructing element $\Id \in S$, the $\mathbb{T}^6/S_4$ orbifold 
contains the following inequivalent constructing elements $g \in S$ from the various twisted sectors
\begin{subequations}
\begin{align}
&g ~=~ \left(\vartheta, 0\right)\;,                                      &&\\
&g ~=~ \left(\omega, \left(n_5 e_5 + n_6 e_6\right)\right)               &\mathrm{for }\ &\ n_5, n_6 \in \{0,1\}\;,\\
&g ~=~ \left(\omega^2, \left(n_5 e_5 + n_6 e_6\right)\right)             &\mathrm{for }\ &\ n_5, n_6 \in \{0,1\}\;,\\
&g ~=~ \left(\omega^2, \left(e_4 + n_5 e_5 + n_6 e_6\right)\right)       &\mathrm{for }\ &\ (n_5, n_6) \in \{(0,1),(1,0),(1,1)\}\;,\\
&g ~=~ \left(\omega^2, \left(e_3 + n_4 e_4 + e_5 + n_6 e_6\right)\right) &\mathrm{for }\ &\ (n_4, n_6) \in \{(0,0),(0,1),(1,1)\}\;,\\
&g ~=~ \left(\vartheta \omega, \left(n_3 e_3 + n_4 e_4\right)\right)     &\mathrm{for }\ &\ n_3, n_4 \in \{0,1\}\;.
\end{align}
\end{subequations}
These $1+4+(4+3+3)+4$ constructing elements are the $S_4$ analogue of the four constructing 
elements of the $\mathbb{T}^2/\Z{2}$ orbifold listed in eq.~\eqref{eq:ConstructingElementsZ2}.

In order to identify the relations on the gauge embeddings between $R_\vartheta$, $R_\omega$ and 
$R_{e_i}$ we consider the action of the twists $(\vartheta,0)$ and $(\omega,0)$ on the basis vectors 
$(\Id,e_i)$ explicitly and embed these relations into $R_g$. Thus, we obtain the conditions
\begin{subequations}\label{eq:S4ConditionThetaE}
\begin{eqnarray}
(\vartheta,0)\, (\Id, e_1)\, (\vartheta^{-1},0) ~=~ (\Id,\phantom{-}e_5) & \Rightarrow & R_\vartheta\, R_{e_1} ~=~ R_{e_5}\,R_\vartheta \;,\label{eq:S4ConditionThetaE1}\\
(\vartheta,0)\, (\Id, e_2)\, (\vartheta^{-1},0) ~=~ (\Id,\phantom{-}e_6) & \Rightarrow & R_\vartheta\, R_{e_2} ~=~ R_{e_6}\,R_\vartheta \;,\label{eq:S4ConditionThetaE2}\\
(\vartheta,0)\, (\Id, e_3)\, (\vartheta^{-1},0) ~=~ (\Id,\phantom{-}e_1) & \Rightarrow & R_\vartheta\, R_{e_3} ~=~ R_{e_1}\,R_\vartheta \;,\label{eq:S4ConditionThetaE3}\\
(\vartheta,0)\, (\Id, e_4)\, (\vartheta^{-1},0) ~=~ (\Id,\phantom{-}e_2) & \Rightarrow & R_\vartheta\, R_{e_4} ~=~ R_{e_2}\,R_\vartheta \;,\label{eq:S4ConditionThetaE4}\\
(\vartheta,0)\, (\Id, e_5)\, (\vartheta^{-1},0) ~=~ (\Id,\phantom{-}e_3) & \Rightarrow & R_\vartheta\, R_{e_5} ~=~ R_{e_3}\,R_\vartheta \;,\label{eq:S4ConditionThetaE5}\\
(\vartheta,0)\, (\Id, e_6)\, (\vartheta^{-1},0) ~=~ (\Id,\phantom{-}e_4) & \Rightarrow & R_\vartheta\, R_{e_6} ~=~ R_{e_4}\,R_\vartheta \;,\label{eq:S4ConditionThetaE6}
\end{eqnarray}
\end{subequations}
and
\begin{subequations}\label{eq:S4ConditionOmegaE}
\begin{eqnarray}
(\omega,0)\, (\Id, e_1)\, (\omega^{-1},0) ~=~ (\Id,\phantom{-}e_1) & \Rightarrow & R_\omega\, R_{e_1} ~=~ R_{e_1}\,R_\omega \;,\label{eq:S4ConditionOmegaE1}\\
(\omega,0)\, (\Id, e_2)\, (\omega^{-1},0) ~=~ (\Id,\phantom{-}e_2) & \Rightarrow & R_\omega\, R_{e_2} ~=~ R_{e_2}\,R_\omega \;,\label{eq:S4ConditionOmegaE2}\\
(\omega,0)\, (\Id, e_3)\, (\omega^{-1},0) ~=~ (\Id,-e_5)           & \Rightarrow & R_\omega\, R_{e_3} ~=~ R_{e_5}^{-1}\,R_\omega \;,\label{eq:S4ConditionOmegaE3}\\
(\omega,0)\, (\Id, e_4)\, (\omega^{-1},0) ~=~ (\Id,-e_6)           & \Rightarrow & R_\omega\, R_{e_4} ~=~ R_{e_6}^{-1}\,R_\omega \;,\label{eq:S4ConditionOmegaE4}\\
(\omega,0)\, (\Id, e_5)\, (\omega^{-1},0) ~=~ (\Id,\phantom{-}e_3) & \Rightarrow & R_\omega\, R_{e_5} ~=~ R_{e_3}\,R_\omega \;,\label{eq:S4ConditionOmegaE5}\\
(\omega,0)\, (\Id, e_6)\, (\omega^{-1},0) ~=~ (\Id,\phantom{-}e_4) & \Rightarrow & R_\omega\, R_{e_6} ~=~ R_{e_4}\,R_\omega \;.\label{eq:S4ConditionOmegaE6}
\end{eqnarray}
\end{subequations}

Let us assume that we have found two matrices $R_{e_1}$ and $R_{e_2}$ that commute 
with $R_\omega$ and, hence, eqs.~\eqref{eq:S4ConditionOmegaE1} and~\eqref{eq:S4ConditionOmegaE2} 
are satisfied. Then, we can solve 
eqs.~\eqref{eq:S4ConditionThetaE1},~\eqref{eq:S4ConditionThetaE2},~\eqref{eq:S4ConditionThetaE3} 
and~\eqref{eq:S4ConditionThetaE4} by defining
\begin{subequations}\label{eq:S4DefE3E4E5E6}
\begin{eqnarray}
R_{e_3} & = & R_\vartheta^{-1}\, R_{e_1} \,R_\vartheta \;,\\
R_{e_4} & = & R_\vartheta^{-1}\, R_{e_2} \,R_\vartheta \;,\\
R_{e_5} & = & R_\vartheta\, R_{e_1} \,R_\vartheta^{-1} \;,\\
R_{e_6} & = & R_\vartheta\, R_{e_2} \,R_\vartheta^{-1} \;.
\end{eqnarray}
\end{subequations}

This choice automatically satisfies eqs.~\eqref{eq:S4ConditionThetaE5} 
and~\eqref{eq:S4ConditionThetaE6} using $(R_\vartheta)^3 = \Id$. Consequently, we are left with the 
conditions~\eqref{eq:S4ConditionOmegaE3},~\eqref{eq:S4ConditionOmegaE4},~\eqref{eq:S4ConditionOmegaE5} 
and~\eqref{eq:S4ConditionOmegaE6}. Let us start with eqs.~\eqref{eq:S4ConditionOmegaE3} 
and~\eqref{eq:S4ConditionOmegaE5}, i.e.\ we have to demand
\begin{subequations}
\begin{eqnarray}
R_\omega\, R_{e_3} & = & R_{e_5}^{-1}\,R_\omega\;,\\
R_\omega\, R_{e_5} & = & R_{e_3}\,R_\omega\;.
\end{eqnarray}
\end{subequations}
Using the definitions of $R_{e_3}$ and $R_{e_5}$ from eq.~\eqref{eq:S4DefE3E4E5E6} we see that this 
is equivalent to
\begin{subequations}
\begin{eqnarray}
\left(R_\vartheta^2\, R_\omega\, R_\vartheta^2\right)\, R_{e_1} & = & R_{e_1}^{-1}\,\left(R_\vartheta^2\, R_\omega\, R_\vartheta^2\right)\;,\\
\left(R_\vartheta\,   R_\omega\, R_\vartheta\right)\,   R_{e_1} & = & R_{e_1}\,     \left(R_\vartheta\,   R_\omega\, R_\vartheta\right)\;,\label{eq:S4TrivialConditionE1}
\end{eqnarray}
\end{subequations}
Since $R_\vartheta\,   R_\omega\, R_\vartheta = R_{\vartheta\,\omega\,\vartheta} = R_{\omega^{-1}} = (R_{\omega})^{-1}$ 
condition~\eqref{eq:S4TrivialConditionE1} is trivially satisfied using our assumption $R_{e_1} R_{\omega}=R_{\omega} R_{e_1}$. 
As a remark, we see that $R_\vartheta^2\, R_\omega\, R_\vartheta^2 = R_{\vartheta^2\, \omega\, \vartheta^2}$ and 
$(\vartheta^2\, \omega\, \vartheta^2)^2 = \Id$, thus 
\begin{equation}
\left(R_{\vartheta^2\, \omega\, \vartheta^2}\right)^2 ~=~ \Id\;.
\end{equation}

Now, we repeat these steps for eqs.~\eqref{eq:S4ConditionOmegaE4} and~\eqref{eq:S4ConditionOmegaE6} 
and obtain
\begin{subequations}
\begin{eqnarray}
\left(R_\vartheta^2\, R_\omega\, R_\vartheta^2\right)\, R_{e_2} & = & R_{e_2}^{-1}\,\left(R_\vartheta^2\, R_\omega\, R_\vartheta^2\right)\;,\\
\left(R_\vartheta\,   R_\omega\, R_\vartheta\right)\,   R_{e_2} & = & R_{e_2}\,     \left(R_\vartheta\,   R_\omega\, R_\vartheta\right)\;.\label{eq:S4TrivialConditionE2}
\end{eqnarray}
\end{subequations}
Again, using $R_\vartheta\,   R_\omega\, R_\vartheta = (R_{\omega})^{-1}$ and $R_{e_2} R_{\omega}=R_{\omega} R_{e_2}$ we see 
that condition~\eqref{eq:S4TrivialConditionE2} is trivial.

\bibliography{Orbifold}
\bibliographystyle{OurBibTeX}

\end{document}